\documentclass[a4paper,11pt]{article} 
\usepackage{jcappub} 

\usepackage[utf8]{inputenc} 
\usepackage{amsfonts,amssymb, mathrsfs, amsmath, esint,bm,enumitem}
\allowdisplaybreaks

\usepackage{subcaption}
\usepackage{ragged2e}
\DeclareCaptionJustification{justified}{\justifying}
\usepackage{diagbox}
\usepackage{latexsym}
\usepackage{graphicx}
\usepackage[dvipsnames]{xcolor}
\usepackage{booktabs,multirow}
\usepackage{titlesec} 
\usepackage{physics}
\usepackage[normalem]{ulem}
\graphicspath{{./Figures/}}

\usepackage{tikz}
\usetikzlibrary{positioning,calc}

\usepackage{hyperref}
\hypersetup{colorlinks, citecolor=ForestGreen, linkcolor=BrickRed, urlcolor=RedViolet}

\usepackage[capitalize]{cleveref}
\usepackage{float}

\newcommand{\AIC}{\Delta \textrm{AIC}}

\newcommand{\out}[1]{\textcolor{OliveGreen}{\sout{#1}}}

\renewcommand{\out}[1]{}

\newcommand{\planck}{\textbf{P18}}
\newcommand{\planckhilli}{\textbf{P20\_H}}
\newcommand{\desi}{$\mathbf{+}$\textbf{DESI}}
\newcommand{\boss}{$\mathbf{+}$\textbf{SDSS+6dFGS}}
\newcommand{\pantheon}{$\mathbf{+}${\bf Pantheon\_Plus}}
\newcommand{\bbnlike}{$\mathbf{+ Y_\text{He}, D/H}$}
\newcommand{\shoes}{$\mathbf{+ H_0}$}
\newcommand{\DES}{$\mathbf{+}$\textbf{DES-SN5YR}}

\newcommand{\DNeff}{\Delta N_\text{eff}}

\newcommand{\MCMC}{\textrm{\tiny{MC}}}
\newcommand{\beq}{\begin{equation}}
\newcommand{\eeq}{\end{equation}}

\title{Reduced Hubble Tension in Dark Radiation Models after DESI 2024
}

\author[a]{Itamar J. Allali}
\author[b,c]{Alessio Notari}
\author[d,e]{Fabrizio Rompineve}

\affiliation[a]{ Department of Physics, Brown University, Providence, RI 02912, USA\looseness=-1}
\affiliation[b]{Departament de F\'isica Qu\`antica i Astrofis\'ica \& Institut de Ci\`encies del Cosmos (ICCUB), Universitat de Barcelona, Mart\'i i Franqu\`es 1, 08028 Barcelona, Spain. 		\looseness=-1}
\affiliation[c]{ Galileo Galilei Institute for theoretical physics, Centro Nazionale INFN di Studi Avanzati Largo Enrico Fermi 2, I-50125, Firenze, Italy		\looseness=-1}
\affiliation[d]{Departament de F\'isica, Universitat Aut\`onoma de Barcelona, 08193 Bellaterra, Barcelona, Spain}
\affiliation[e]{Institut de F\'isica d’Altes Energies (IFAE) and The Barcelona Institute of Science and Technology (BIST), Campus UAB, 08193 Bellaterra (Barcelona), Spain}

\emailAdd{itamar\_allali@brown.edu}
\emailAdd{notari@fqa.ub.edu}
\emailAdd{frompineve@ifae.es}

\abstract{
We investigate the presence of extra relativistic degrees of freedom in the early Universe, contributing to the effective number of neutrinos $N_\text{eff}$, as $\Delta N_\text{eff}\equiv N_\text{eff}-3.044\geq 0$, in light of the recent measurements of Baryon Acoustic Oscillations (BAO) by the DESI collaboration. We analyze one-parameter extensions of the $\Lambda$CDM model where dark radiation (DR) is free streaming or behaves as a perfect fluid, due to self-interactions. We report a significant relaxation of upper bounds on $\Delta N_\text{eff}$, with respect to previous BAO data from SDSS+6dFGS, when additionally employing \emph{Planck} data (and supernovae data from \emph{Pantheon+}), setting $\Delta N_\text{eff}\leq 0.39$ ($95\%$ C.L.) for free streaming DR, and a very mild preference for fluid DR, $\Delta N_\text{eff} = 0.221^{+0.088}_{-0.18}$ ($\leq 0.46$, $95\%$ C.L.).  
Applying constraints from primordial element abundances leads to tighter constraints on $\DNeff$, but they
are avoided if DR is produced after Big Bang Nucleosynthesis (BBN).
For fluid DR we estimate the tension with the SH$_0$ES determination of $H_0$ to be less than $3\sigma$ and as low as $2\sigma$,
and for free-streaming DR the tension is below $3\sigma$ if production occurs after BBN. This lesser degree of tension motivates a combination with SH$_0$ES in these cases, 
resulting in a $4.4\sigma-5\sigma$ evidence for dark radiation with $\Delta N_\text{eff}\simeq 0.6$ and large improvements in $\chi^2$ over $\Lambda$CDM, $-18\lesssim \Delta \chi^2\lesssim -25$. Upcoming data releases by DESI and other CMB and LSS surveys will decisively confirm or disfavour this conclusion. 
}

\begin{document}
\maketitle
\vspace{1em}\noindent

\section{Introduction}
\label{sec:introduction}

Modern cosmological datasets are among the most powerful probes of physics beyond the Standard Model (SM), even when this has negligible interactions with SM particles. This is particularly true if new physics is in the form of light degrees of freedom that remain ultra-relativistic throughout the cosmological evolution, until after the epoch of recombination. Their additional contribution to the energy density impacts the background expansion and density perturbations in the early Universe, when the Cosmic Microwave Background (CMB) is produced~ (see~\cite{Bashinsky:2003tk, Hou:2011ec} and~\cite{Baumann:2015rya}). Finding evidence for such \emph{dark radiation} (DR), or alternatively constraining its presence to unprecedented levels, is one of the main targets of active and future cosmological surveys~\cite{ACT:2020gnv,SPT-3G:2021vps,Amendola:2016saw,LSST:2008ijt,CMB-S4:2022ght,SimonsObservatory:2018koc}, and has a potentially groundbreaking impact on fundamental physics.

The aim of this work is to assess the status of DR in light of the new measurements of Baryon Acoustic Oscillations (BAO) from galaxies and quasars~\cite{DESI:2024uvr} at redshifts $0.3 \lesssim z\lesssim 1.5$ and from the Lyman-$\alpha$ forest~\cite{DESI:2024lzq} by the Dark Energy Spectroscopic Instrument (DESI).

BAO data from previous galactic surveys~\cite{Beutler:2011hx, Ross:2014qpa, BOSS:2016ntk} have so far provided the most stringent constraints on DR, when combined with CMB measurements from the Planck satellite~\cite{Planck:2018vyg} (BBN and measurements of primordial element abundances provide an alternative probe, though one with possibly larger uncertainties, see e.g.~\cite{Planck:2019nip}, and~\cite{Yeh:2022heq}). In terms of the customary parameterization of the abundance of DR, given by the \emph{effective number of neutrino species}, i.e. $\Delta N_\text{eff}\equiv \rho_\text{DR}/\rho_\nu$, where $\rho_\nu$ is the energy density of a single neutrino species  {in the instantaneous decoupling limit}, the DESI collaboration has recently reported $\Delta N_\text{eff}\leq 0.40$ ($95\%~\text{C.L.}$)~\cite{DESI:2024mwx} for free streaming species. Interestingly, this is a significant relaxation of the previous CMB+BAO bound $\Delta N_\text{eff}\leq 0.28$~\cite{Planck:2018vyg} ($95\%~\text{C.L.}$, with fixed sum of neutrino masses $\sum m_\nu =0.06~\text{eV}$). Both these results were obtained allowing for $\Delta N_\text{eff}<0$ in the prior.

While the $\Delta N_\text{eff}$ parameterization effectively captures a vast landscape of particle physics scenarios, the specific microphysical origin of DR can lead to different imprints on cosmological observables. Perhaps the simplest model dependence arises between the case where DR is free streaming (some well motivated examples are: the QCD axion with a small mass~\cite{Turner:1986tb, Kolb:1990vq, Berezhiani:1992rk, Chang:1993gm, Masso:2002np, Ferreira:2020bpb, Notari:2022ffe, DEramo:2018vss, Ferreira:2018vjj}, relic gravitational waves,  {sterile neutrinos, see e.g.~\cite{Archidiacono:2022ich},} see also~\cite{Baumann:2016wac} for other candidates), and the possibility that it behaves as a perfect fluid with equation of state parameter $w=1/3$ (see e.g. the discussion in~\cite{Baumann:2015rya}). This latter case applies to a self-interacting gas of relativistic particles (as can arise e.g. in dark sector models with gauge interactions~\cite{Chacko:2015noa, Buen-Abad:2015ova, Chacko:2016kgg, Nakagawa:2022knn}), see~\cite{Cyr-Racine:2015ihg,Lesgourgues:2015wza, Brust:2017nmv, Buen-Abad:2017gxg,Archidiacono:2019wdp, Blinov:2020hmc, Brinckmann:2022ajr} for investigations with previous data, and to scalar fields that start oscillating in quartic potentials well before recombination. 

The first aim of this work is thus to provide the state-of-the-art constraints on such simplest DR scenarios, also accounting for data from additional cosmological observations, such as measurements of the Hubble diagram from supernovae~\cite{Scolnic:2021amr} and of primordial element abundances. These can then be used by particle physicists to determine bounds on microphysical models.

Our findings then lead to the second aim of our work. We indeed interestingly find that the new BAO data allow for larger abundances of DR in all cases of study, which motivates a reassessment of whether such simple one parameter extensions of the $\Lambda$CDM model can reconcile the value of the Hubble expansion parameter $H_0$ inferred from fitting to cosmological datasets, with the larger value measured from supernovae~\cite{Riess:2021jrx} (see also \cite{Wong:2019kwg,Scolnic:2023mrv, Freedman:2021ahq} for other measurements).

\section{Models and datasets}\label{sec:model}

 {We limit our study to two simple realizations of DR, that have exactly the same background evolution, and differ only at the level of perturbations. In particular we consider DR that is either \emph{free-streaming}, with large anisotropic stress~\cite{Bashinsky:2003tk} that produces a phase shift in CMB anisotropies; or \emph{fluid-like}, i.e. with vanishing anisotropic stress and thus standard Euler and continuity equations. Both models are effectively captured by a single parameter $\Delta N_\text{eff}\equiv N_\text{eff}-3.044$, where the latter contribution comes from SM neutrinos~\cite{Froustey:2020mcq, Akita:2020szl, Bennett:2020zkv, Drewes:2024wbw}. It is well known that CMB bounds on $\Delta N_\text{eff}$ are tighter for free-streaming than for fluid DR~\cite{Baumann:2015rya, Blinov:2020hmc}. While strictly speaking both models involve additional massless relics, they effectively capture any scenario where the mass of DR particles is too small to be probed by cosmological data.}

Throughout our work, we take neutrinos to be degenerate in mass and temperature and impose the prior $\sum m_\nu\geq 0.06~\text{eV}$ from neutrino oscillations.\footnote{Using a less informed prior $\sum m_\nu\geq 0$ does not significantly alter the results on dark radiation, although the resulting posteriors on neutrino masses exhibit better agreement with the bound from neutrino oscillations than in the $\Lambda$CDM model~\cite{Allali:2024aiv}.} We therefore always add $\sum m_\nu$ as an additional free cosmological parameter to the $\Lambda$CDM model. We consider scenarios where the neutrino sector is not altered with respect to the SM prediction, studying only additional radiation degrees of freedom, which justifies the prior $\Delta N_\text{eff}\geq 0$.~\footnote{This differs from the choice of the DESI collaboration, the prior choice of which allows neutrinos to be colder than as predicted by the SM.  {We investigate in \cref{app:posteriors} the possibility that the neutrino abundance is altered after BBN.}}

We perform Bayesian searches using {\tt CLASS}~\cite{Lesgourgues:2011re, Blas:2011rf} to solve for the cosmological evolution and {\tt MontePython}~\cite{Audren:2012wb, Brinckmann:2018cvx} to collect Markov Chain Monte Carlo (MCMC) samples. We obtain posteriors and figures using {\tt GetDist}~\cite{Lewis:2019xzd}. We consider the following datasets in our searches:  Planck 2018 high-$\ell$ and low-$\ell$ TT, TE, EE and lensing data~\cite{Planck:2019nip} (\planck); BAO measurements from DESI 2024~\cite{DESI:2024mwx} (\desi); Previous BAO measurements from 6dFGS~\cite{Beutler:2011hx} and SDSS~\cite{Ross:2014qpa, BOSS:2016wmc}, which we use only in alternative to DESI data (\boss); the Pantheon+ supernovae compilation~\cite{Scolnic:2021amr} as implemented in {\tt MontePython} by the likelihood {\tt Pantheon\_Plus} (\pantheon);  {and} measurements of the abundances of Helium~\cite{Aver:2021rwi} and Deuterium (see~\cite{Yeh:2022heq} and refs. therein). The theoretical predictions for $Y_\text{He}, D/H$ at BBN are obtained according to the default likelihood {\tt bbn} of {\tt MontePython} (\bbnlike).

For the purposes of setting constraints, we will consider the combination \planck\desi \\\pantheon\, to be our baseline dataset. Weaker bounds from \planck\desi\, alone are reported in \cref{app:planckdesi}. The \bbnlike\, dataset is used to generate constraints when appropriate.

\begin{table*}
\centering
\begin{tabular} {| l | c| c| c| c|}
\hline\hline
 \multicolumn{1}{|c|}{ Parameter}&   \multicolumn{2}{|c|}{\planck\desi\pantheon} &  \multicolumn{2}{|c|}{\bbnlike}\\
 \hline
  \multicolumn{1}{|c|}{}&  \multicolumn{1}{|c|}{~Free-streaming~} &  \multicolumn{1}{|c|}{~~~Fluid~~~}  &\multicolumn{1}{|c|}{~Free-streaming~} &  \multicolumn{1}{|c|}{~~~Fluid~~~}
  \\
\hline\hline
$\Delta N_{\mbox{eff}}$  &  $ < 0.386$  & $0.221^{+0.088}_{-0.18}$ & {$ < 0.248$}  & {$ < 0.304$}
\\
\hline
$H_0 \,[\mathrm{km}/\mathrm{s}/\mathrm{Mpc}]$ &$68.79^{+0.60}_{-0.89}     $ & $69.35^{+0.81}_{-1.1}      $  & {$68.66^{+0.49}_{-0.65}     $} & {$68.96^{+0.58}_{-0.82}     $}
\\
\hline\hline
$H_0$ tension &$3.06\sigma $ & $2.52\sigma $ &{$3.47\sigma $} & {$3.11\sigma $}
\\
\hline
\end{tabular}
\caption{Marginalized posteriors for $\Delta N_\text{eff}$ and $H_0$. Two models for dark radiation are considered: free-streaming and perfect fluid. We report results with our baseline dataset, and additionally adding measurements of primordial abundances. We report upper bounds on $\Delta N_\text{eff}$·at $95\%$ C.L. for all models and datasets, except for the fluid model fitted to the baseline dataset, where a $1\sigma$ preference for dark radiation is found (the $95\%$ C.L. upper bound is $\DNeff < 0.461$). The corresponding tension with the SH$_0$ES measurement is also reported. Posteriors for all parameters are reported in \cref{app:posteriors}.}
\label{tab:constraints}
\end{table*}


\section{New constraints on dark radiation}\label{sec:constraints}

\begin{figure}
    \centering
    \includegraphics[width=0.55\textwidth]{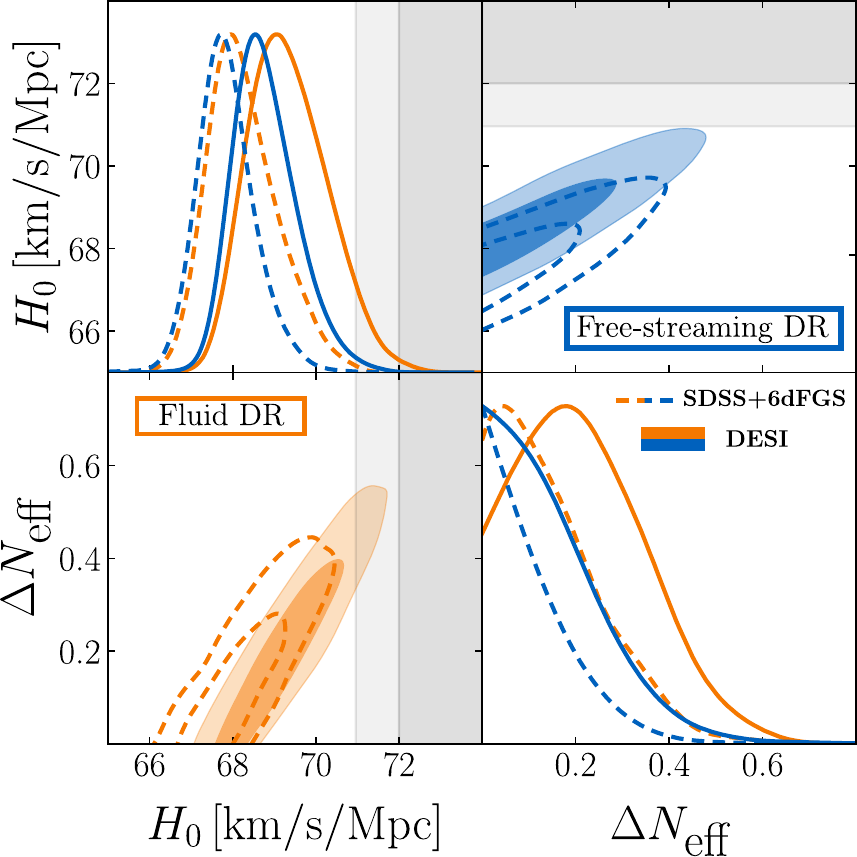}
    \caption{1- and 2-d posterior distributions for $H_0$ and $\DNeff$ in dark radiation models, obtained using our baseline dataset. We compare our results with the new DESI BAO data (solid curves/shaded contours) with those obtained with previous BAO data (dashed curves/contours). The 68\% (95\%) confidence intervals from the measurement of $H_0$ by SH$_0$ES are shown in the (lighter) gray shaded region.}
    \label{fig:bossvdesi}
\end{figure}

Let us first focus on the impact of the new BAO data from DESI on DR models. We fit the models of free-streaming and fluid DR to the baseline dataset, and compare this to the case where previous BAO data are used instead of DESI BAO. Posteriors for $\Delta N_\text{eff}$ and $H_0$ are shown in \cref{fig:bossvdesi}. One can immediately appreciate the qualitative difference in the results; for both free-streaming and fluid DR, the $1$ and $2\sigma$ regions of the posteriors extend to larger values of $\DNeff$, indicating that the DESI BAO data allow for larger abundances of DR. The new  $95\% \,\text{C.L.}$ constraints are reported in Table~\ref{tab:constraints}, and show a significant relaxation 
of up to $20\%$ for free-streaming DR with respect to using \boss, see also \cref{app:posteriors} (our $95\%\, \text{C.L.}$ upper bound on $\Delta N_\text{eff}$ agrees with~\cite{DESI:2024mwx}, despite our different prior choice; the central value is however shifted to larger values than in~\cite{DESI:2024mwx}, as expected).  {The 1d marginalized posterior for $\Delta N_\text{eff}$ in fluid DR is shifted to even larger values, and a non vanishing abundance $\Delta N_\text{eff}\approx 0.2$ is now (very mildly) preferred at $1\sigma$ (although this preference is reduced in the 2d posterior).} In all our runs we find similar upper bounds $\sum m_\nu\leq 0.12\sim 0.13~\text{eV}$. Further posteriors for all models and cosmological parameters are reported in in Table~\ref{tab:constraints} and \cref{app:posteriors}.

The DR abundance allowed by the new BAO data is potentially independently constrained by observations of light element abundances. We therefore examine the impact of these measurements using the \bbnlike\, dataset. We report the resulting upper bounds on $\Delta N_\text{eff}$ in \cref{tab:constraints}. One can see that the mild preference for $\DNeff>0$ in the fluid case is erased by the \bbnlike\, data. Constraints on $\DNeff$ become significantly tighter for the free-streaming case as well.

\begin{figure}[t]
    \centering
    \includegraphics[width=0.95\textwidth]{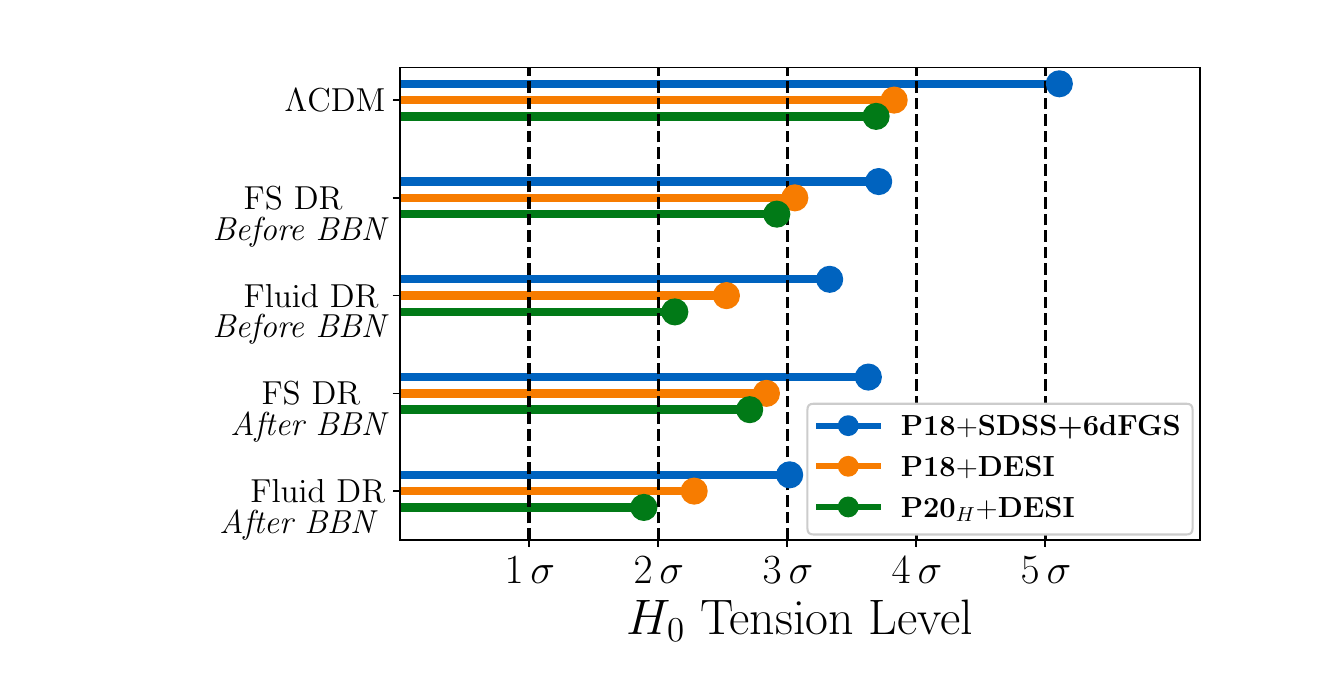}
    \caption{Measure of the tension (IT, see \cref{app:integrated_tension}) in the determinations of $H_0$ from the \planck\desi\pantheon~and \planckhilli\desi\pantheon \,  datasets with respect to the SH$_0$ES measurement, for models considered in this work. Results with previous BAO data are shown for comparison. FS refers to ``free-streaming."}
    \label{fig:lollipop}
\end{figure}

We now move to $H_0$, and highlight two interesting effects of the new BAO data: first, larger values are preferred, even in the absence of dark radiation. Second, the alleviated constraints on $\Delta N_\text{eff}$ allow for even larger values of $H_0$, given the strong degeneracy between these two parameters. For comparison, the SH$_0$ES measurement~\cite{Riess:2021jrx} is shown by the gray shaded bands ($1-$ and $2\sigma$) in Fig.~\ref{fig:bossvdesi}. We report estimates of the Hubble tension in Table~\ref{tab:constraints}, as computed by using the true posterior distribution from our MCMC analysis (rather than the simpler Gaussian estimates, see~\cite{Raveri:2021wfz} and \cref{app:integrated_tension} for a review). We estimate the tension to be around $2.5\sigma$ within the fluid model, and around $3\sigma$ in the free-streaming scenario, when we do not include constraints from primordial elements (the tension is further lowered by $\sim 0.3\sigma$ using \planck\desi~alone, see \cref{app:planckdesi}). These results represent a significant alleviation of the $H_0$ tension within these models. Comparing to results obtained with previous BAO data, we find that DESI reduces the tension by $(0.5-1)\sigma$ depending on the model, see Fig.~\ref{fig:lollipop}.  {The DR models considered here provide fits to cosmological datasets as good as the $\Lambda$CDM model (as assessed by their $\Delta\chi^2$, see  {the tables in \cref{app:posteriors}}).} 

\section{The Hubble tension}\label{sec:Hubble tension}

Additional constraints from primordial elements worsen the tension in all models. However, they are avoided if the DR is produced sufficiently after the epoch of BBN and before recombination. This specification does not introduce any additional parameters, nor does it lead to a coincidence problem (in contrast to models where a fluid is taken to undergo a transition around the epoch of recombination, such as~\cite{Poulin:2018cxd, Niedermann:2019olb, Gonzalez:2020fdy, Allali:2021azp, Aloni:2021eaq, Allali:2023zbi}), since DR can still be produced in a wide redshift range, corresponding to $\text{eV}\ll T\ll 100~\text{keV}$. The limiting cases of DR production close to BBN or to recombination are potentially interesting, although they require a detailed modeling and thus additional parameters. Our results do not apply to these scenarios; we leave their investigation to future work.

 {Furthermore, when DR is present during BBN, it alters} the theoretical prediction of $Y_\text{He}$, which then affects the number density of electrons at recombination $n_e(z)\propto (1-Y_\text{He})$ (see e.g.~\cite{Cyr-Racine:2021oal} for a recent discussion). Therefore, when considering the scenario where DR is produced after BBN, we determine $Y_\text{He}$ by setting $\Delta N_\text{eff}=0$ at BBN. 
We compare the inferences made with our baseline dataset for both fluid and free-streaming DR produced after BBN in \cref{tab:plusshoes}.
Interestingly, we find an additional reduction of the $H_0$ tension (and relaxation of constraints on $\DNeff$), which now stands at $(2.3-2.8)\sigma$, and a $\gtrsim 1\sigma$ preference for fluid DR persists.

\begin{figure}
    \centering
    \includegraphics[width=0.48\textwidth]{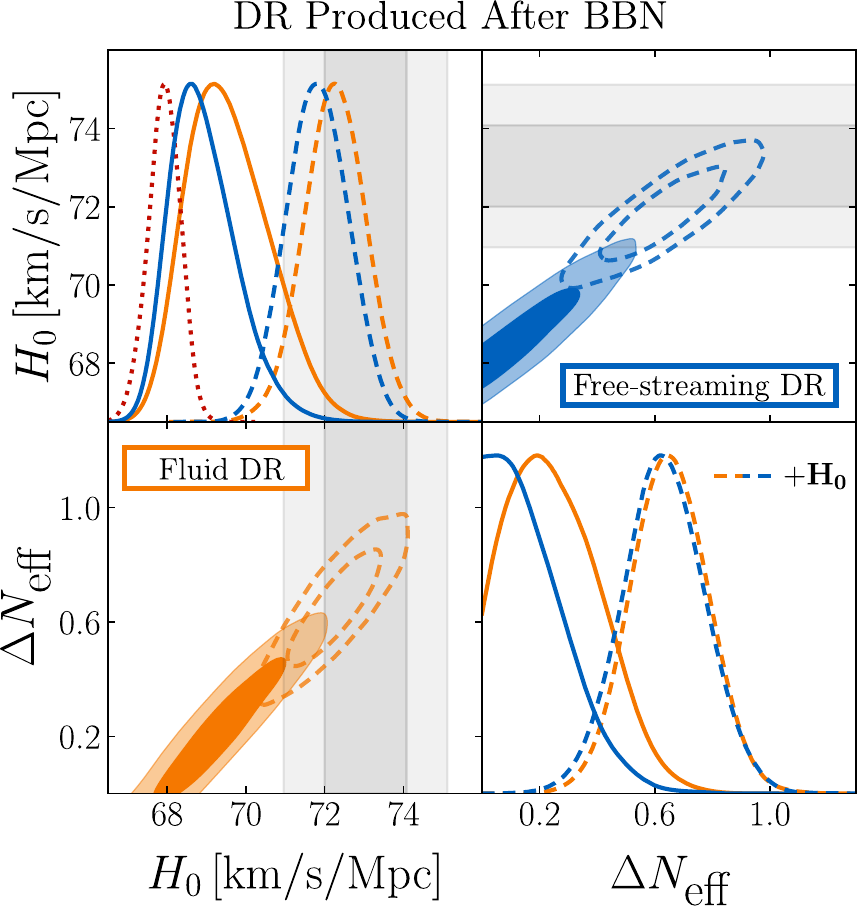}
    \includegraphics[width=0.48\textwidth]{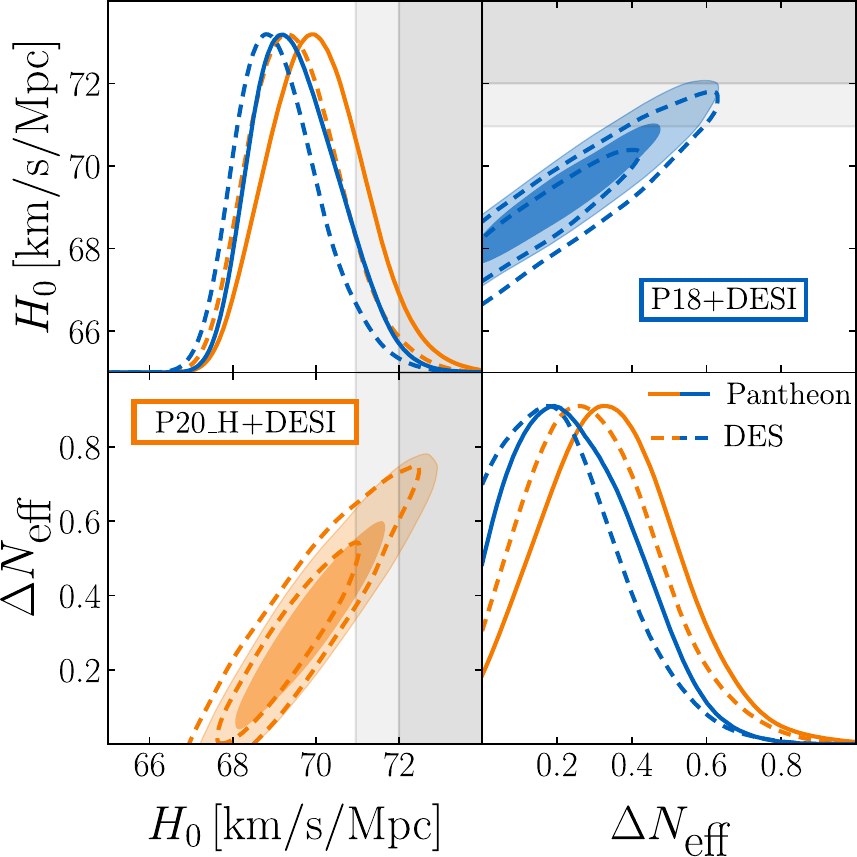}
    \caption{\textbf{\textit{Left:}} 1- and 2-d posterior distributions for $H_0$ and $\DNeff$ in models with dark radiation produced after BBN. We compare our results obtained with our baseline dataset  (solid curves/shaded contours) with those obtained by combining with the determination of $H_0$ from SH$_0$ES (dashed curves/contours). The 1-d $H_0$ posterior for $\Lambda$CDM with the baseline dataset is shown by the dotted curve. \textbf{\textit{Right:}} 1- and 2-d posterior distributions for $H_0$ and $\DNeff$ for fluid DR produced after BBN with different datasets.}
    \label{fig:SHOES}
\end{figure}

 {This opens the possibility to interpret the Hubble tension as a mild statistical fluctuation within the context of the DR models with production after BBN. We thus combine our baseline dataset with the measurement of the intrinsic SNIa magnitude $M_b=-19.253\pm0.027$ from the SH$_{0}$ES collaboration~\cite{Riess:2021jrx}, as consistently implemented with Pantheon+ data in the {\tt Pantheon\_Plus\_SHOES} likelihood in {\tt MontePython} (\shoes).} Results are presented in the rightmost columns of \cref{tab:plusshoes}  
 and the left side of \cref{fig:SHOES}.
 Remarkably, we find evidence at the $5\sigma$ ($4.5\sigma$) level for fluid (free-streaming) DR, and a negligible residual tension with SH$_0$ES. This is accompanied by a very significant improvement in $\chi^2$ with respect to the $\Lambda$CDM model. We account for the additional parameter via the Akaike Information Criterion (AIC)~\cite{1100705} (see also~\cite{Liddle:2007fy}) $\AIC\equiv \Delta\chi^2 + 2\times($\# of added free parameters$)$ and report $\AIC\simeq -23(-19)$ for fluid (free-streaming) DR.

{Let us now investigate the impact of alternative CMB and supernovae likelihoods/data on these conclusions. We focus on the fluid DR model} and present results using the Planck PR4 Hillipop/Lollipop likelihoods~\cite{Tristram:2023haj} (\planckhilli) and the DES-SN5YR supernovae dataset ~\cite{DES:2024tys} in 
the right side of \cref{fig:SHOES}
and \cref{tab:newdata}. With \planckhilli, which has displayed a resolution to the anomalous amplitude of the lensing power spectrum $A_L$ in previous Planck likelihoods \cite{Addison:2023fqc}, we find even further evidence for $\DNeff$ and a reduced $H_0$ tension to $\sim 1.9 \sigma$ (when combining with \pantheon). When replacing Pantheon+ with DES-SN5YR, the tension is only marginally worsened by at most $0.3\sigma$. A quantitatively similar effect occurs when using the latest value of $H_0$ from SH$_0$ES~\cite{Breuval:2024lsv}.  {The tension thus remains below $2.8\sigma$. The use of alternative CMB likelihoods~\cite{Rosenberg:2022sdy} and/or supernovae data~\cite{Rubin:2023ovl} is not expected to significantly impact these conclusions.}

In all the models and combination of datasets used in this work, we find posteriors for the matter clustering parameter $S_8\equiv \sigma_8\sqrt{\Omega_m/0.3}$  {compatible at better than $1.5\sigma$} with weak lensing surveys~\cite{Kilo-DegreeSurvey:2023gfr}, see \cref{app:posteriors}.

Finally, we have focused here on perhaps the simplest DR scenarios, but it is conceivable that analogous conclusions could apply to other scenarios that lead to a similar evolution for the cosmological background, for instance models with varying Newton constant, see e.g.~\cite{Ballesteros:2020sik}.

\begin{table*}
\centering
\resizebox{1.0\textwidth}{!}{
\begin{tabular} {| l | c| c| c| c|}
\hline\hline
 \multicolumn{1}{|c|}{}&   \multicolumn{2}{|c|}{\planck\desi\pantheon} & \multicolumn{2}{|c|}{\shoes}\\
 \hline
 \multicolumn{1}{|c|}{ Parameter} & \multicolumn{1}{|c|}{Free-streaming DR} &  \multicolumn{1}{|c|}{~~~~Fluid DR~~~~} &   \multicolumn{1}{|c|}{Free-streaming DR} &  \multicolumn{1}{|c|}{~~~~Fluid DR~~~~}\\
\hline\hline
$\Delta N_{\mbox{eff}}$    & $ < 0.435$ & $0.26~(0.34)^{+0.11}_{-0.21}      $ & $0.63~(0.56)\pm 0.14      $ & $0.65~(0.73)\pm 0.13      $\\
$H_0 \,[\mathrm{km}/\mathrm{s}/\mathrm{Mpc}]$  &$68.94~(68.41)^{+0.63}_{-0.99}     $ & $69.56~(69.82)^{+0.85}_{-1.2}      $ &$71.82~(71.65)^{+0.78}_{-0.77}     $ & $72.26~(73.0)^{+0.77}_{-0.78}     $\\
\hline
\hline
$H_0$ tension & $2.84\sigma $ & $2.28\sigma $ & $0.94\sigma $ & $0.6\sigma $\\
\hline
$\Delta \chi^2$ & $\sim 0$ & $-0.4$ & $-20.5$ & $-24.7$\\
\hline
$\Delta$AIC & $+2.0$ & $+1.6$ & $-18.5$ & $-22.7$\\
\hline
\end{tabular}
}
\caption{Marginalized posteriors for $\Delta N_\text{eff}$ and $H_0$ for scenarios where dark radiation is produced after BBN  {(upper bounds are reported at $95\%$ C.L., bestfit values in parentheses)}. Two models are considered: free-streaming and fluid DR. We report results with our baseline dataset, and additionally adding the determination of $M_b$ from SH$_0$ES. The corresponding tension with the SH$_0$ES measurement is also reported, as well as two measures of goodness-of-fit compared to the $\Lambda$CDM model. Posteriors for all parameters are reported in \cref{app:posteriors}.}
\label{tab:plusshoes}

\end{table*}

\begin{table*}
\centering
\resizebox{1.0\textwidth}{!}{
 \begin{tabular} {| l| c| c| c| c|}
\hline\hline
\multicolumn{1}{|c|}{}&   \multicolumn{2}{|c|}{~~\planckhilli\desi~~} &  \multicolumn{2}{|c|}{\planck\desi}\\
 \hline
 \multicolumn{1}{|c|}{ Parameter} & \multicolumn{1}{|c|}{\pantheon} &  \multicolumn{1}{|c|}{\DES} &  \multicolumn{1}{|c|}{\pantheon} &  \multicolumn{1}{|c|}{\DES}\\
\hline\hline
$\Delta N_{\mbox{eff}}$   & $0.35~(0.18)^{+0.15}_{-0.20}      $ & $0.31~(0.31)^{+0.13}_{-0.21}      $ & $0.26~(0.34)^{+0.11}_{-0.21}      $ & $0.231~(0.019)^{+0.062}_{-0.22}    $\\
$H_0 \,[\mathrm{km}/\mathrm{s}/\mathrm{Mpc}]$ & $70.0~(69.4)^{+1.0}_{-1.3}        $ & $69.54~(69.78)^{+0.92}_{-1.2}      $ & $69.56~(69.82)^{+0.85}_{-1.2}      $ & $69.13~(68.07)^{+0.79}_{-1.2}      $\\
\hline
\hline
$H_0$ tension & $1.87\sigma $ & $2.21\sigma $ & $2.28\sigma $ & $2.51\sigma $\\
\hline
\end{tabular}
}
\caption{Marginalized posteriors for $\Delta N_\text{eff}$ and $H_0$ for fluid dark radiation produced after BBN with different datasets.}
\label{tab:newdata}

\end{table*}



\section{Discussion}\label{sec:discussions}

The large improvements in the goodness-of-the-fit which we find are of course driven by the SH$_0$ES measurement, and have been reported to a similar level in the past for other models with previous BAO data. Therefore, one may doubt the relevance of our results. However, we would like to stress two crucial differences with such previous findings. First, our results are obtained by combining datasets that are in $\sim 2.5\sigma$ tension with each other within the context of DR models when production occurs only after BBN, especially for the fluid DR case (removing Pantheon+ data or using the PR4 likelihood further reduces the tension to even $\sim2\sigma$, see \cref{app:posteriors} and \cref{fig:lollipop}, respectively). Second, the models under consideration are simple, one-parameter extensions of $\Lambda$CDM, with several possible implementations in particle physics. 

Perhaps the simplest example of DR production after BBN, which can arise in a broad class of models, is a massive particle, the abundance of which is negligible in the BBN era, and that decays to some light states after BBN (see e.g.~\cite{Ichikawa:2007jv, Fischler:2010xz, Hooper:2011aj, Bjaelde:2012wi, Choi:2012zna, Hasenkamp:2012ii, Sobotka:2023bzr}),
with a decay rate 
$10^{-37}~\text{GeV}\ll\Gamma\ll 10^{-27}~\text{GeV}$.
Other possibilities exist in the literature for the production of DR after BBN; see, for instance, \cite{Ferreira:2022zzo, Aloni:2023tff, Garny:2024ums}. Any such model, as long as the species decays efficiently only into DR, can be captured effectively with the one-parameter extension (adding $\DNeff$) we consider in this work, as long as the decay occurs sufficiently after BBN and before recombination.

Our work provides new state-of-the-art bounds on dark radiation models, that partially relax constraints on beyond the SM physics and should prove important for model building.

Remarkably, our findings also suggest that the new BAO data open the possibility to address the Hubble tension with well-motivated minimal extensions of $\Lambda$CDM model. 

With a conservative perspective, the possibility that our findings are driven by a statistical fluctuation or underestimated systematic uncertainties in the DESI measurement (which exhibit some discrepancy with previous BAO data, see~\cite{DESI:2024mwx}) should be kept in mind, and will be decisively clarified soon by upcoming data releases from DESI itself and Euclid, see e.g.~\cite{Brinckmann:2018owf}, where the $1\sigma$ error on $\Delta N_\text{eff}$ for the free streaming case from Euclid power spectrum and lensing measurements is forecasted to be $0.05$.


\begin{acknowledgments}
We thank H\'ector Gil-Mar\'in  for help with DESI data, and Maria Vincenzi and Dillon Brout for help with the DES-SN5YR likelihood. The work of F.R.~is supported by the grant RYC2021-031105-I from the Ministerio de Ciencia e Innovación (Spain). I.J.A. is supported by NASA grant 80NSSC22K081. The work of A.N. is supported by the grants PID2019-108122GB-C32 from the Spanish Ministry of Science and Innovation, Unit of Excellence Maria de Maeztu 2020-2023 of ICCUB (CEX2019-000918-M) and AGAUR 2021 SGR 00872. A.N. is grateful to the Physics Department of the University of Florence for the hospitality during the course of this work.
We acknowledge use of the Tufts HPC research cluster and the INFN Florence cluster.
\end{acknowledgments}


\bibliographystyle{JHEP}
\bibliography{biblio.bib}

\appendix 


\section{Tension Measures}\label{app:integrated_tension}

For the assessment of tension between the $H_0$ measurement of the SH$_0$ES collaboration and the inferences made in this work, we define the following metrics. First, the commonly used measure of ``Gaussian tension" ($GT$) is defined as 

\beq
GT = \frac{|\mu_m-\mu_\MCMC|}{\sqrt{\sigma_m^2+\sigma_\MCMC^2}},
\label{GT}
\eeq
where $\mu_m$ and $\mu_\MCMC$ are the mean values of $H_0$ determined by the SH$_0$ES collaboration and by our MCMC analyses, respectively. $\sigma_m^2$ is the variance for the SH$_0$ES measurement. For the variance of the MCMC inference $\sigma_\MCMC^2$, we take the upper $1\sigma$ error derived from our marginalized posteriors on $H_0$. Since the posteriors for $H_0$ we derive are not symmetric, there is not a clear choice of whether to average the upper and lower $\sigma$, or to take the value that is on the side of the distribution closest to the SH$_0$ES measurement (upper). In this work, we take $\sigma_\MCMC$ to always be the upper derived $\sigma$ such that we do not underestimate the tension. 

To address the non-gaussian nature of our inferred posteriors, we employ also the measure which we term the ``integrated tension" ($IT$) \cite{Allali:2023zbi,Raveri:2021wfz}, defined via

\beq
\int^{\infty}_{-\infty} \mathcal{P}_{\MCMC}(h)\frac{1}{2}\left(1 \pm \mbox{erf}\left(\frac{h-\mu_{m}}{\sqrt{2}\sigma_{m}}\right)\right) dh = \int_{-\infty}^{IT}  \frac{1}{\sqrt{2\pi}}e^{-\frac{1}{2}x^2}dx
\label{eq:IT}
\eeq

To understand this formula, let us examine each side. The left hand side is the integral over the cross-correlation of the two posterior distributions, namely the posterior distribution derived from our MCMC $\mathcal{P}_{\MCMC}(h)$, and the posterior from the SH$_0$ES measurement. In \cref{eq:IT}, we have already integrated the SH$_0$ES posterior, assuming it to be purely gaussian, and therefore all that remains are the mean and standard deviation $\mu_m$ and $\sigma_m$. Using the posterior distribution from a given MCMC, the left hand side constitutes a probability (understood as the probability of measuring the SH$_0$ES value given the posterior from MCMC). Then, the right hand side of \cref{eq:IT} equates this probability with the integral over a gaussian (with mean$=0$ and variance$=1$), and one solves for the upper limit of the integral $IT$ which gives this same probability. For example, if the left hand side of \cref{eq:IT} gives a probability of $68\%$, then we obtain the measure of tension to be $IT=1\sigma$.

\section{Modified neutrino abundance and Detailed Posteriors}\label{app:posteriors}

We present in the following sections the detailed posteriors we obtain when evaluating the several models discussed in this work against several combinations of datasets.

 {Before moving to the presentation of the posteriors for the DR models presented in the main text, let us mention a different simple scenario that is captured by the $\Delta N_\text{eff}$ parametrization: one where the abundance of neutrinos differs from the prediction of the SM, as can be the case if neutrinos or photons are slightly reheated by a dark sector after their decoupling (i.e. at temperatures below MeV). Notice that this scenario differs from the DR models considered in the main text both at the background and perturbation levels, since SM neutrinos have a non-negligible mass around and after recombination. We have investigated this scenario by allowing for the SM neutrino temperature to differ from that predicted by the SM (both larger or smaller), while keeping the free-streaming nature of neutrinos and we present the corresponding results in this section, together with the posteriors for the DR models. Overall we find that the neutrino scenario is less relevant for the Hubble tension than the DR models, although deviations of up to $6\%$ from the SM abundance are allowed by the baseline dataset.}

\cref{app:planckdesi,app:planckdesipantheon,app:planckdesipantheonshoes,app:planckdesibbnlikepantheon,app:planckbosspantheon,app:planckbosspantheon} present tables and plots of posteriors for the models: $\Lambda$CDM, Free-streaming dark radiation, Fluid dark radiation, and Neutrinos. The title of each section gives the combination of data explored in that section. Then, \cref{app:FLD_afterbbn} and \cref{app:FS_afterbbn} explore the posteriors with a variety of datasets on the models of dark radiation produced after BBN for fluid and free-streaming, respectively.
 {Finally, \cref{app:planckhilli_DES} presents tables and plots of posteriors for the Fluid DR model, with the DR produced after BBN, including the \planckhilli~ and \DES~ datasets. Within the context of \planckhilli\pantheon, we have checked the effects of changing the BAO likelihood. Using the BAO measurements from 6dFGS at $z = 0.106$~\cite{Beutler:2011hx}; SDSS MGS at $z = 0.15$~\cite{Ross:2014qpa}; and SDSS eBOSS DR16 measurements \cite{eBOSS:2020yzd}, we find $\DNeff = 0.198^{+0.058}_{-0.19}$. Using instead the combination suggested in~\cite{DESI:2024mwx} that merges DESI 2024 with previous SDSS measurements, we find $\DNeff=0.219^{+0.088}_{-0.18}$.} In the following, the acronyms GT and IT stay for ``Gaussian Tension" and ``Ìntegrated Tension" respectively, corresponding to different measures of the Hubble tension. In the main text we have reported the IT.

\pagebreak
\subsection{\planck\desi}\label{app:planckdesi}
\begin{table}[H]
\centering
\resizebox{1.0\textwidth}{!}{
\begin{tabular} {| l | c| c| c| c|}
\hline\hline
 \multicolumn{1}{|c|}{ Parameter} &  \multicolumn{1}{|c|}{$\Lambda$CDM} &  \multicolumn{1}{|c|}{Free-streaming DR} &  \multicolumn{1}{|c|}{Fluid DR} &  \multicolumn{1}{|c|}{Neutrinos}\\
\hline\hline
$100 \omega_b$             & $2.249~(2.248)^{+0.013}_{-0.013}   $ & $2.262~(2.264)^{+0.016}_{-0.016}   $ & $2.272~(2.274)^{+0.017}_{-0.019}   $ & $2.256~(2.25)^{+0.019}_{-0.017}   $\\
$\omega_{cdm }             $ & $0.11811~(0.11823)^{+0.00087}_{-0.00086}$ & $0.1208~(0.1212)^{+0.0015}_{-0.0024}$ & $0.1224~(0.1205)^{+0.0021}_{-0.0031}$ & $0.1193~(0.1176)^{+0.0024}_{-0.0028}$\\
$\ln 10^{10}A_s$           & $3.054~(3.057)^{+0.014}_{-0.016}   $ & $3.062~(3.055)^{+0.015}_{-0.017}   $ & $3.052~(3.041)^{+0.015}_{-0.016}   $ & $3.058~(3.057)^{+0.015}_{-0.018}   $\\
$n_{s }                    $ & $0.9689~(0.9689)^{+0.0036}_{-0.0036}$ & $0.9748~(0.9742)^{+0.0047}_{-0.0057}$ & $0.9712~(0.9678)^{+0.0039}_{-0.0039}$ & $0.9713~(0.9715)^{+0.0064}_{-0.0065}$\\
$\tau_{reio }              $ & $0.0608~(0.0608)^{+0.0070}_{-0.0081}$ & $0.0614~(0.0595)^{+0.0072}_{-0.0083}$ & $0.0619~(0.053)^{+0.0072}_{-0.0084}$ & $0.0615~(0.0641)^{+0.0067}_{-0.0087}$\\
$\Delta N_{\mbox{eff}}$    & -- & $ < 0.395$ & $0.25~(0.13)^{+0.11}_{-0.18}      $ & $0.12~(0.062)^{+0.16}_{-0.16}      $\\
$\sum m_\nu$               & $< 0.119                  $ & $< 0.124                 $ & $< 0.127                  $ & $< 0.116                  $\\
\hline
$H_0 \,[\mathrm{km}/\mathrm{s}/\mathrm{Mpc}]$ & $68.09~(68.18)^{+0.43}_{-0.40}     $ & $69.10~(68.91)^{+0.66}_{-0.95}     $ & $69.75~(69.19)^{+0.87}_{-1.2}      $ & $68.5~(68.4)^{+1.1}_{-0.99}       $\\
$S_8$                      & $0.813~(0.818)^{+0.010}_{-0.010}   $ & $0.814~(0.817)^{+0.011}_{-0.011}   $ & $0.812~(0.809)^{+0.010}_{-0.010}   $ & $0.816~(0.807)^{+0.011}_{-0.010}   $\\
\hline
$H_0$ GT & $4.4\sigma $ & $3.2\sigma $ & $2.43\sigma $ & $3.0\sigma $\\
\hline
$H_0$ IT & $4.12\sigma $ & $2.81\sigma $ & $2.17\sigma $ & $3.08\sigma $\\
\hline
\end{tabular}
}
\caption{Marginalized posteriors for various model parameters for the $\Lambda$CDM, Free-streaming DR, Fluid DR, and Neutrino models, fitting to the dataset: \planck\desi. All upper bounds are reported at 95\% C.L., for any case where the $1\sigma$ lower bound is overlapping with our priors.}
\end{table}

\begin{figure}[H]
\centering
    \includegraphics[width=0.7\textwidth]{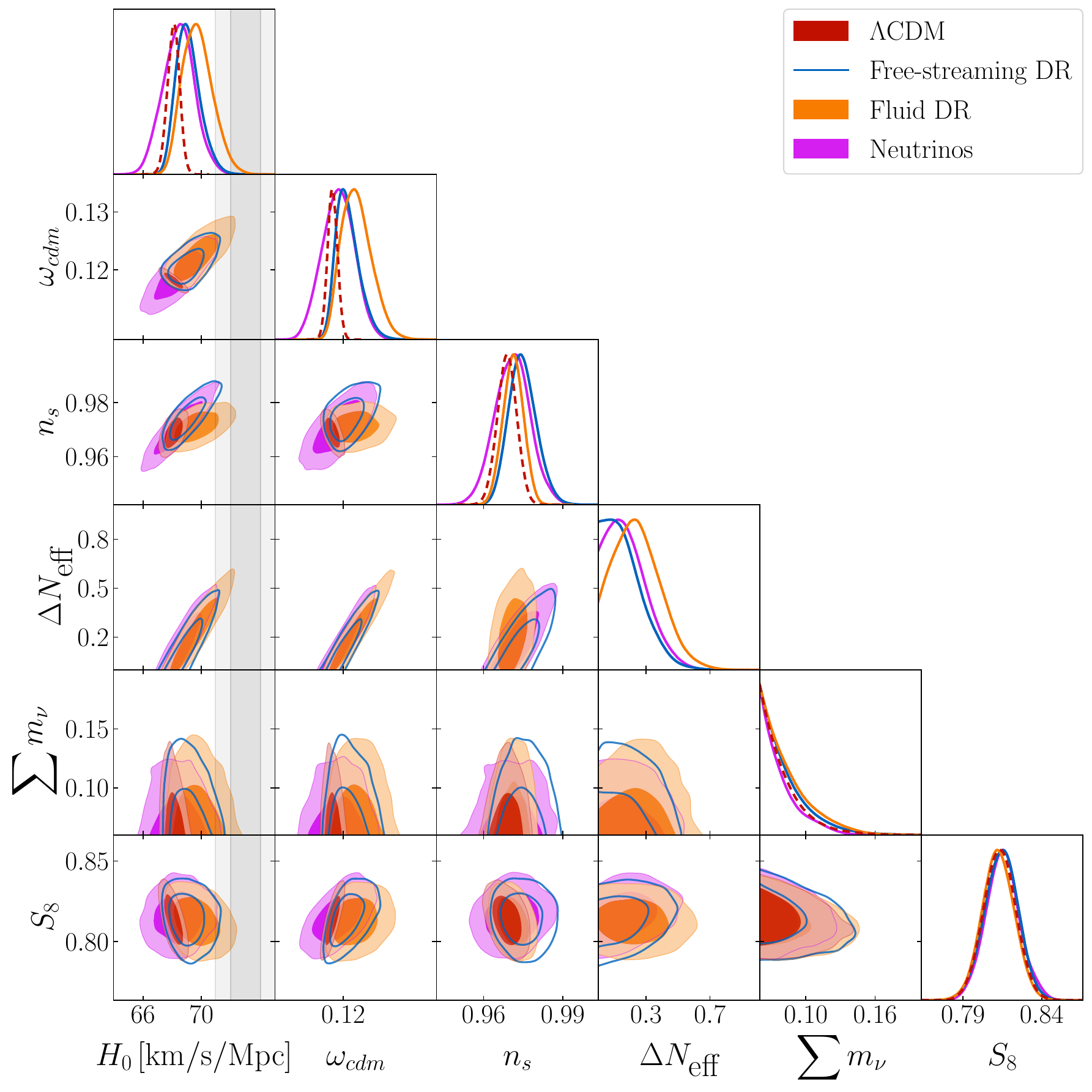}
    \caption{One and two-dimensional posterior distributions for various model parameters for the $\Lambda$CDM, Free-streaming DR, Fluid DR, and Neutrino models, fitting to the dataset: \planck\desi. {The 68\% and 95\% confidence intervals from the measurement of $H_0$ by SH$_0$ES are shown in the gray and lighter gray shaded regions.}}
\end{figure}

\begin{table*}
\centering
\begin{tabular} {| l | c| c| c| c|}
\hline\hline
Dataset & $\Lambda$CDM & Free-streaming DR & Fluid DR & Neutrinos\\
\hline
Planck\_highl\_TTTEEE &  $2353.71$ &  $+0.56$ &  $+3.19$ &  $+1.39$\\
Planck\_lowl\_EE &  $397.62$ &  $-0.46$ &  $-1.82$ &  $+0.29$\\
Planck\_lowl\_TT &  $22.82$ &  $-0.59$ &  $+0.25$ &  $-0.92$\\
Planck\_lensing &  $9.0$ &  $+0.17$ &  $+0.41$ &  $+0.61$\\
DESI\_BAO &  $15.76$ &  $+0.70$ &  $-1.02$ &  $-1.17$\\
DESI\_BAO\_DV &  $1.14$ &  $+0.14$ &  $-0.32$ &  $-0.37$\\
Total &  $2800.05$ &  $+0.53$ &  $+0.68$ &  $-0.16$\\
\hline\hline
\end{tabular}
\caption{Values of $\chi^2$ for each likelihood when fit to a combination of \planck\desi, reported as the difference from $\Lambda$CDM for the other models.}
\end{table*}

\subsection{\planck\desi\pantheon}\label{app:planckdesipantheon}
\begin{table}[H]
\centering
\resizebox{1.0\textwidth}{!}{
\begin{tabular} {| l | c| c| c| c|}
\hline\hline
  \multicolumn{1}{|c|}{ Parameter} &  \multicolumn{1}{|c|}{$\Lambda$CDM} &  \multicolumn{1}{|c|}{Free-streaming DR} &  \multicolumn{1}{|c|}{Fluid DR} &  \multicolumn{1}{|c|}{Neutrinos}\\
\hline\hline
$100 \omega_b$             & $2.247~(2.251)^{+0.013}_{-0.013}   $ & $2.258~(2.246)^{+0.015}_{-0.016}   $ & $2.265~(2.257)^{+0.017}_{-0.019}   $ & $2.248~(2.256)^{+0.019}_{-0.018}   $\\
$\omega_{cdm }             $ & $0.11844~(0.11856)^{+0.00084}_{-0.00086}$ & $0.1211~(0.1192)^{+0.0014}_{-0.0025}$ & $0.1223~(0.1212)^{+0.0020}_{-0.0030}$ & $0.1186~(0.1191)^{+0.0028}_{-0.0031}$\\
$\ln 10^{10}A_s$           & $3.054~(3.061)^{+0.015}_{-0.016}   $ & $3.061~(3.039)^{+0.014}_{-0.017}   $ & $3.050~(3.055)^{+0.015}_{-0.015}   $ & $3.054~(3.052)^{+0.017}_{-0.018}   $\\
$n_{s }                    $ & $0.9681~(0.9679)^{+0.0039}_{-0.0036}$ & $0.9734~(0.9689)^{+0.0045}_{-0.0058}$ & $0.9699~(0.9667)^{+0.0037}_{-0.0039}$ & $0.9685~(0.9759)^{+0.0068}_{-0.0069}$\\
$\tau_{reio }              $ & $0.0602~(0.0636)^{+0.0074}_{-0.0083}$ & $0.0606~(0.0537)^{+0.0071}_{-0.0082}$ & $0.0605~(0.0619)^{+0.0072}_{-0.0080}$ & $0.0601~(0.0608)^{+0.0071}_{-0.0086}$\\
$\Delta N_{\mbox{eff}}$    & -- & $ < 0.386$ & $0.221~(0.128)^{+0.088}_{-0.18}    $ & $0.06~(0.143)^{+0.17}_{-0.19}      $\\
$\sum m_\nu$               & $< 0.123                  $ & $< 0.137                  $ & $< 0.132                  $ & $< 0.127                $\\
\hline
$H_0 \,[\mathrm{km}/\mathrm{s}/\mathrm{Mpc}]$ & $67.93~(68.07)^{+0.44}_{-0.38}     $ & $68.79~(67.99)^{+0.60}_{-0.89}     $ & $69.35~(68.72)^{+0.81}_{-1.1}      $ & $68.0~(67.14)^{+0.97}_{-1.2}      $\\
$S_8$                      & $0.817~(0.822)^{+0.010}_{-0.010}   $ & $0.818~(0.813)^{+0.010}_{-0.011}   $ & $0.8161~(0.825)^{+0.0099}_{-0.0099}$ & $0.817~(0.821)^{+0.011}_{-0.011}   $\\
$M_b$                      & $-19.424~(-19.421)^{+0.013}_{-0.011} $ & $-19.396~(-19.426)^{+0.017}_{-0.028} $ & $-19.381~(-19.4)^{+0.024}_{-0.033} $ & $-19.422~(-19.448)^{+0.030}_{-0.036} $\\
\hline
$H_0$ GT & $4.53\sigma $ & $3.53\sigma $ & $2.81\sigma $ & $3.54\sigma $\\
\hline
$H_0$ IT & $3.93\sigma $ & $3.06\sigma $ & $2.52\sigma $ & $3.22\sigma $\\
\hline
\end{tabular}
}
\caption{Marginalized posteriors for various model parameters for the $\Lambda$CDM, Free-streaming DR, Fluid DR, and Neutrino models, fitting to the dataset: \planck\desi\pantheon. All upper bounds are reported at 95\% C.L., for any case where the $1\sigma$ lower bound is overlapping with our priors.}
\end{table}

\begin{figure}[H]
\centering
    \includegraphics[width=0.7\textwidth]{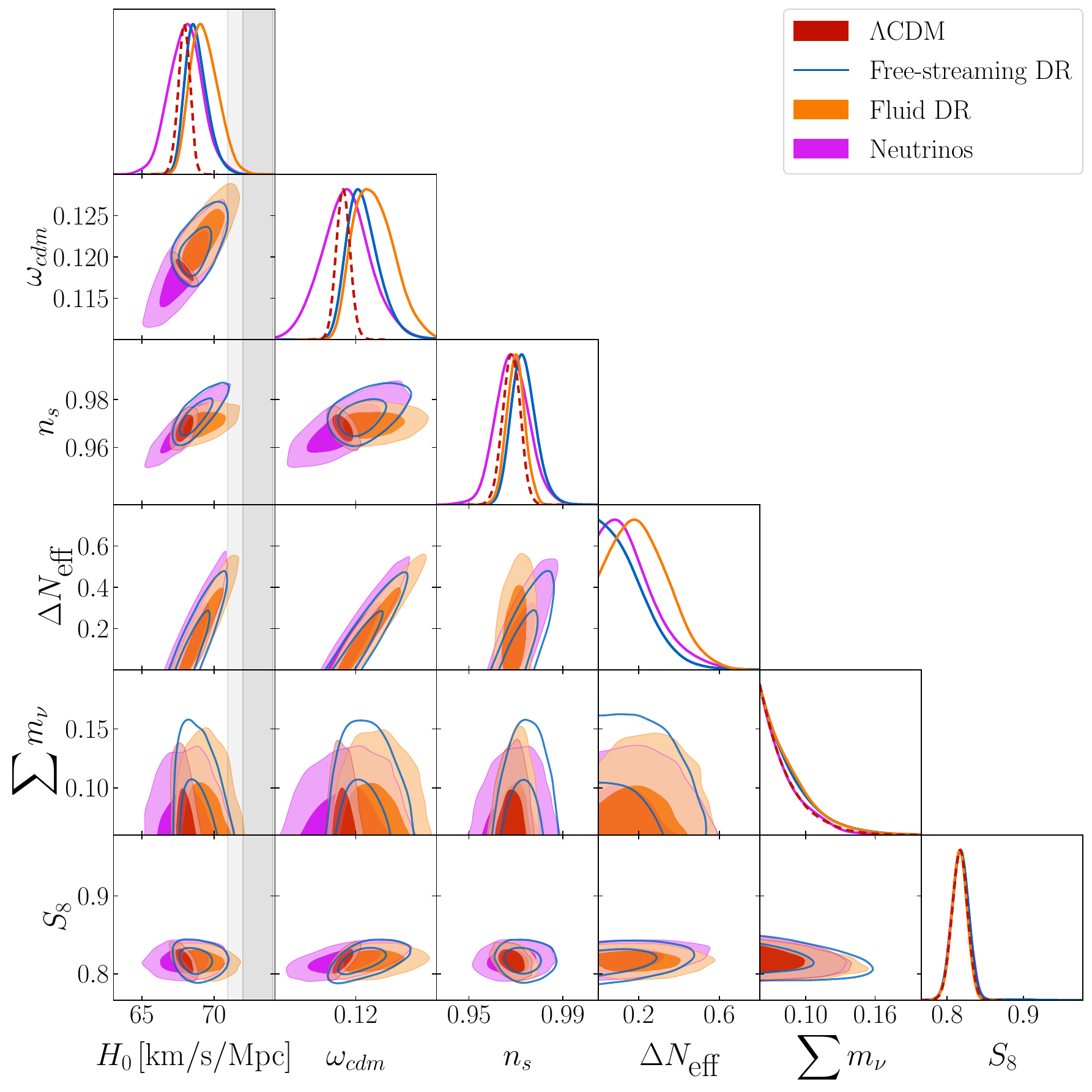}
    \caption{One and two-dimensional posterior distributions for various model parameters for the $\Lambda$CDM, Free-streaming DR, Fluid DR, and Neutrino models, fitting to the dataset: \planck\desi\pantheon. {The 68\% and 95\% confidence intervals from the measurement of $H_0$ by SH$_0$ES are shown in the gray and lighter gray shaded regions.}}
\end{figure}

\begin{table*}
\centering
\begin{tabular} {| l | c| c| c| c|}
\hline\hline
Dataset & $\Lambda$CDM & Free-streaming DR & Fluid DR & Neutrinos\\
\hline
Planck\_highl\_TTTEEE &  $2352.87$ &  $+2.45$ &  $+0.44$ &  $+0.71$\\
Planck\_lowl\_EE &  $398.78$ &  $-2.90$ &  $-0.55$ &  $+0.78$\\
Planck\_lowl\_TT &  $23.17$ &  $-0.45$ &  $+0.15$ &  $-0.32$\\
Planck\_lensing &  $8.75$ &  $+0.60$ &  $+0.01$ &  $+0.40$\\
Pantheon\_Plus &  $1412.17$ &  $+0.58$ &  $+0.02$ &  $+0.84$\\
DESI\_BAO &  $16.57$ &  $+0.98$ &  $0.00$ &  $-1.57$\\
DESI\_BAO\_DV &  $1.36$ &  $+0.27$ &  $-0.01$ &  $-0.42$\\
Total &  $4213.67$ &  $+1.53$ &  $+0.07$ &  $+0.41$\\
\hline\hline
\end{tabular}

\caption{Values of $\chi^2$ for each likelihood when fit to a combination of \planck\desi\pantheon, reported as the difference from $\Lambda$CDM for the other models.}
\end{table*}

\subsection{\planck\desi\pantheon\shoes}\label{app:planckdesipantheonshoes}
\begin{table}[H]
\centering
\resizebox{1.0\textwidth}{!}{
\begin{tabular} {| l | c| c| c| c|}
\hline\hline
 \multicolumn{1}{|c|}{ Parameter} &  \multicolumn{1}{|c|}{$\Lambda$CDM} &  \multicolumn{1}{|c|}{Free-streaming DR} &  \multicolumn{1}{|c|}{Fluid DR} &  \multicolumn{1}{|c|}{Neutrinos}\\
\hline\hline
$100 \omega_b$             & $2.264~(2.275)^{+0.013}_{-0.013}   $ & $2.290~(2.289)^{+0.015}_{-0.015}   $ & $2.304~(2.307)^{+0.015}_{-0.015}   $ & $2.291~(2.285)^{+0.014}_{-0.014}   $\\
$\omega_{cdm }             $ & $0.11682~(0.11669)^{+0.00083}_{-0.00083}$ & $0.1263~(0.126)^{+0.0025}_{-0.0025}$ & $0.1281~(0.1286)^{+0.0019}_{-0.0016}$ & $0.1263~(0.1268)^{+0.0024}_{-0.0024}$\\
$\ln 10^{10}A_s$           & $3.061~(3.07)^{+0.015}_{-0.016}   $ & $3.078~(3.079)^{+0.015}_{-0.017}   $ & $3.048~(3.042)^{+0.015}_{-0.016}   $ & $3.078~(3.065)^{+0.016}_{-0.016}   $\\
$n_{s }                    $ & $0.9723~(0.9732)^{+0.0037}_{-0.0036}$ & $0.9871~(0.9867)^{+0.0049}_{-0.0050}$ & $0.9746~(0.972)^{+0.0036}_{-0.0039}$ & $0.9872~(0.9873)^{+0.0049}_{-0.0050}$\\
$\tau_{reio }              $ & $0.0651~(0.0666)^{+0.0074}_{-0.0085}$ & $0.0634~(0.0636)^{+0.0072}_{-0.0086}$ & $0.0633~(0.0588)^{+0.0074}_{-0.0077}$ & $0.0633~(0.0567)^{+0.0073}_{-0.0084}$\\
$\Delta N_{\mbox{eff}}$    & -- & $0.54~(0.52)^{+0.13}_{-0.13}      $ & $0.592~(0.611)^{+0.091}_{-0.060}   $ & $0.59~(0.619)^{+0.12}_{-0.13}      $\\
$\sum m_\nu$               & $< 0.099                  $ & $< 0.126                  $ & $< 0.131                  $ & $< 0.118                  $\\
\hline
$H_0 \,[\mathrm{km}/\mathrm{s}/\mathrm{Mpc}]$ & $68.82~(68.98)^{+0.37}_{-0.39}     $ & $71.47~(71.39)^{+0.73}_{-0.76}     $ & $72.13~(72.25)^{+0.61}_{-0.41}     $ & $71.46~(71.79)^{+0.73}_{-0.73}     $\\
$S_8$                      & $0.8017~(0.8045)^{+0.0096}_{-0.010} $ & $0.822~(0.824)^{+0.011}_{-0.011}   $ & $0.8095~(0.8086)^{+0.0097}_{-0.010} $ & $0.821~(0.819)^{+0.011}_{-0.011}   $\\
$M_b$                      & $-19.398~(-19.392)^{+0.011}_{-0.011} $ & $-19.320~(-19.319)^{+0.021}_{-0.021} $ & $-19.301~(-19.295)^{+0.017}_{-0.011} $ & $-19.320~(-19.311)^{+0.021}_{-0.021} $\\
\hline
$H_0$ GT & $3.82\sigma $ & $1.23\sigma $ & $0.75\sigma $ & $1.24\sigma $\\
\hline
$H_0$ IT & $3.8\sigma $ & $1.23\sigma $ & $0.76\sigma $ & $1.24\sigma $\\
\hline
$\Delta \chi^2$ & $-$ & $-19.1$ & $-23.8$ & $-17.5$\\
\hline
$\Delta$AIC & $-$ & $-17.1$ & $-21.8$ & $-15.5$\\
\hline
\end{tabular}
}
\caption{Marginalized posteriors for various model parameters for the $\Lambda$CDM, Free-streaming DR, Fluid DR, and Neutrino models, fitting to the dataset: \planck\desi\pantheon\shoes. All upper bounds are reported at 95\% C.L., for any case where the $1\sigma$ lower bound is overlapping with our priors.}
\end{table}

\begin{figure}[H]
\centering
    \includegraphics[width=0.7\textwidth]{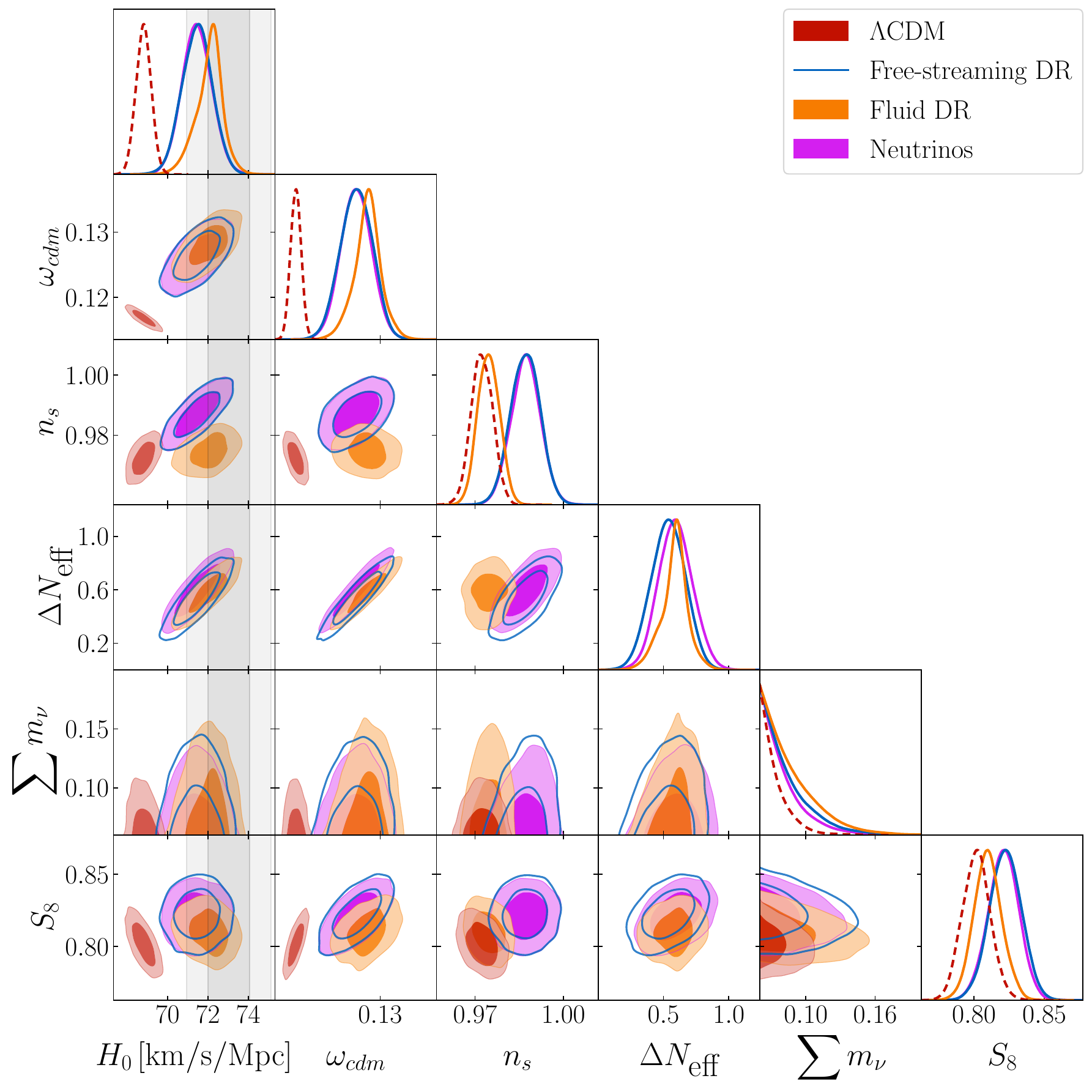}
    \caption{One and two-dimensional posterior distributions for various model parameters for the $\Lambda$CDM, Free-streaming DR, Fluid DR, and Neutrino models, fitting to the dataset: \planck\desi\pantheon\shoes. {The 68\% and 95\% confidence intervals from the measurement of $H_0$ by SH$_0$ES are shown in the gray and lighter gray shaded regions.}}
\end{figure}

\begin{table*}
\centering
\begin{tabular} {| l | c| c| c| c|}
\hline\hline
Dataset & $\Lambda$CDM & Free-streaming DR & Fluid DR & Neutrinos\\
\hline
Planck\_highl\_TTTEEE &  $2360.72$ &  $+3.73$ &  $+1.60$ &  $+8.18$\\
Planck\_lowl\_EE &  $399.7$ &  $-1.35$ &  $-2.78$ &  $-3.40$\\
Planck\_lowl\_TT &  $22.46$ &  $-1.28$ &  $-0.47$ &  $-1.43$\\
Planck\_lensing &  $8.91$ &  $+0.47$ &  $+1.34$ &  $+1.34$\\
Pantheon\_Plus\_shoes &  $1318.58$ &  $-19.79$ &  $-23.57$ &  $-22.16$\\
DESI\_BAO &  $13.54$ &  $+0.17$ &  $-0.15$ &  $-0.10$\\
DESI\_BAO\_DV &  $0.41$ &  $0.00$ &  $-0.10$ &  $-0.10$\\
Total &  $4124.33$ &  $-18.05$ &  $-24.14$ &  $-17.69$\\
\hline\hline
\end{tabular}

\caption{Values of $\chi^2$ for each likelihood when fit to a combination of \planck\desi\pantheon\shoes, reported as the difference from $\Lambda$CDM for the other models.}
\end{table*}

\subsection{\planck\desi\bbnlike\pantheon}\label{app:planckdesibbnlikepantheon}

\begin{table}[H]
\centering
\resizebox{1.0\textwidth}{!}{
\begin{tabular} {| l | c| c| c|}
\hline\hline
 \multicolumn{1}{|c|}{ Parameter} &  \multicolumn{1}{|c|}{$\Lambda$CDM} &  \multicolumn{1}{|c|}{Free-streaming DR} &  \multicolumn{1}{|c|}{Fluid DR} \\
\hline\hline
$100 \omega_b$             & $2.252~(2.25)^{+0.012}_{-0.012}   $ & $2.259~(2.247)^{+0.014}_{-0.014}   $ & $2.263~(2.263)^{+0.015}_{-0.015}   $\\
$\omega{}_{cdm }           $ & $0.11822~(0.11849)^{+0.00081}_{-0.00080}$ & $0.1197~(0.1178)^{+0.0011}_{-0.0016}$ & $0.1205~(0.1195)^{+0.0014}_{-0.0021}$\\
$\ln 10^{10}A_s$           & $3.056~(3.056)^{+0.014}_{-0.016}   $ & $3.059~(3.051)^{+0.014}_{-0.017}   $ & $3.054~(3.056)^{+0.014}_{-0.017}   $\\
$n_{s }                    $ & $0.9700~(0.9687)^{+0.0035}_{-0.0035}$ & $0.9729~(0.9738)^{+0.0040}_{-0.0046}$ & $0.9709~(0.9724)^{+0.0036}_{-0.0036}$\\
$\tau{}_{reio }            $ & $0.0612~(0.0592)^{+0.0071}_{-0.0082}$ & $0.0611~(0.0595)^{+0.0069}_{-0.0083}$ & $0.0616~(0.0637)^{+0.0072}_{-0.0085}$\\
$\Delta N_{\mbox{eff}}$    & -- & $< 0.248                  $ & $< 0.304                   $\\
$\sum m_\nu$               & $< 0.117                  $ & $< 0.132                 $ & $< 0.128                  $\\
\hline
$H_0 \,[\mathrm{km}/\mathrm{s}/\mathrm{Mpc}]$ & $68.11~(68.13)^{+0.38}_{-0.37}     $ & $68.66~(68.38)^{+0.49}_{-0.65}     $ & $68.96~(68.46)^{+0.58}_{-0.82}     $\\
$S_8$                      & $0.8118~(0.816)^{+0.0098}_{-0.0098}$ & $0.813~(0.808)^{+0.011}_{-0.010}   $ & $0.812~(0.817)^{+0.010}_{-0.010}   $\\
$M_b$                      & $-19.419~(-19.416)^{+0.011}_{-0.011} $ & $-19.401~(-19.414)^{+0.014}_{-0.021} $ & $-19.392~(-19.406)^{+0.017}_{-0.026} $\\
\hline
$H_0$ GT & $4.45\sigma $ & $3.81\sigma $ & $3.43\sigma $\\
\hline
$H_0$ IT & $4.34\sigma $ & $3.47\sigma $ & $3.11\sigma $\\
\hline
$\Delta \chi^2$ & $0.0$ & $0.8$ & $0.8$\\
\hline
$\Delta$AIC & $2.0$ & $2.8$ & $2.8$\\
\hline
\end{tabular}
}
\caption{Marginalized posteriors for various model parameters for the $\Lambda$CDM, Free-streaming DR, and Fluid DR models, fitting to the dataset: \planck\desi\bbnlike\pantheon. All upper bounds are reported at 95\% C.L., for any case where the $1\sigma$ lower bound is overlapping with our priors.}
\end{table}

\begin{figure}[H]
\centering
    \includegraphics[width=0.7\textwidth]{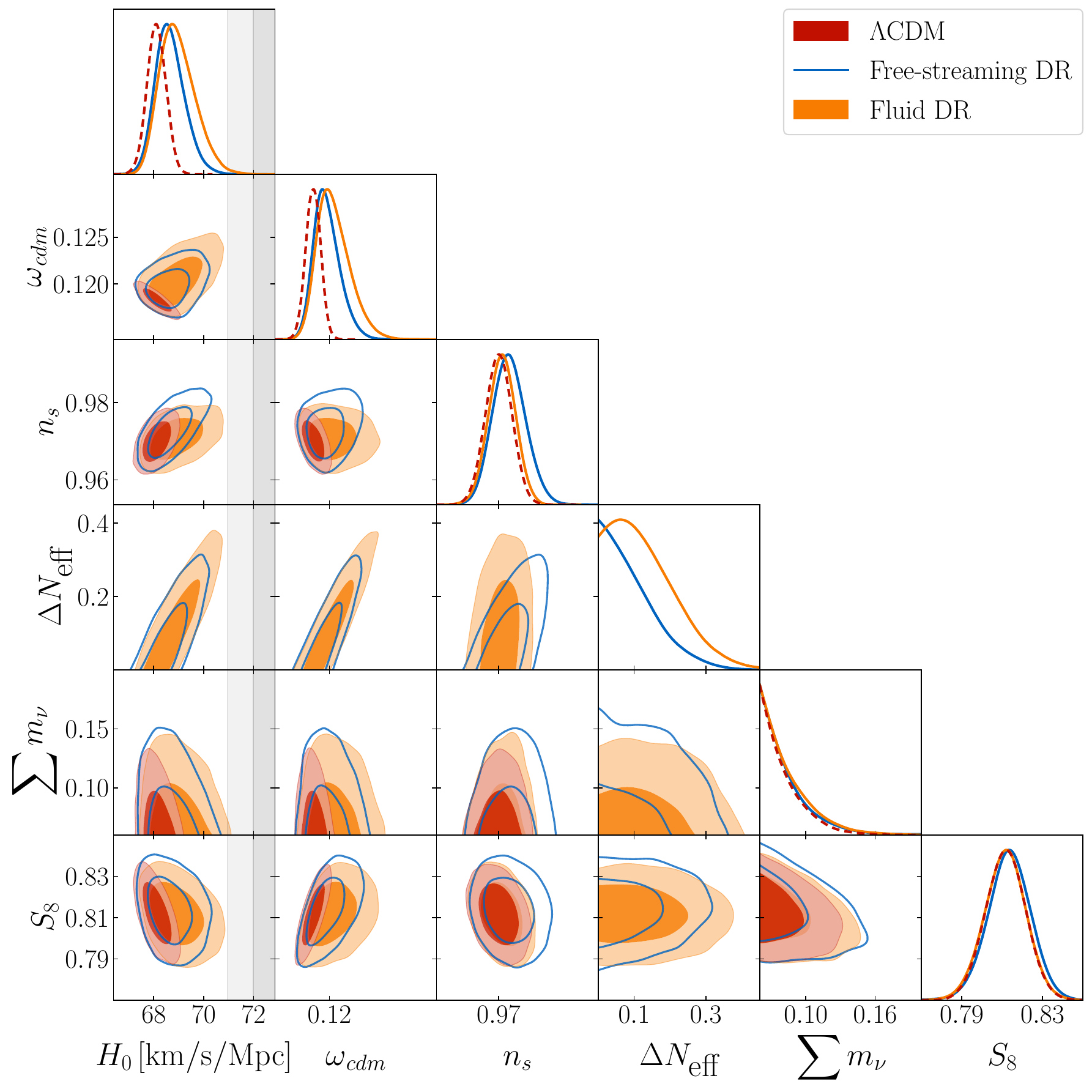}
    \caption{One and two-dimensional posterior distributions for various model parameters for the $\Lambda$CDM, Free-streaming DR, and Fluid DR models, fitting to the dataset: \planck\desi\bbnlike\pantheon. {The 68\% and 95\% confidence intervals from the measurement of $H_0$ by SH$_0$ES are shown in the gray and lighter gray shaded regions.}}
\end{figure}

\begin{table*}
\begin{tabular} {| l | c| c| c|}
\hline\hline
Dataset & $\Lambda$CDM & Free-streaming DR & Fluid DR\\
\hline
Planck\_highl\_TTTEEE &  $2353.49$ &  $+3.52$ &  $+0.68$\\
Planck\_lowl\_EE &  $397.6$ &  $-0.55$ &  $+1.09$\\
Planck\_lowl\_TT &  $23.18$ &  $-1.31$ &  $-0.82$\\
Planck\_lensing &  $8.86$ &  $+0.98$ &  $+0.01$\\
Pantheon\_Plus\_test &  $1412.32$ &  $+0.75$ &  $+0.60$\\
DESI\_BAO &  $16.35$ &  $-1.46$ &  $-0.05$\\
DESI\_BAO\_DV &  $1.29$ &  $-0.40$ &  $-0.01$\\
Total &  $4213.56$ &  $+1.69$ &  $+1.65$\\
\hline\hline
\end{tabular}

\caption{Values of $\chi^2$ for each likelihood when fit to a combination of \planck\desi\bbnlike\pantheon, reported as the difference from $\Lambda$CDM for the other models.}
\end{table*}

\subsection{\planck\boss\pantheon}\label{app:planckbosspantheon}

\begin{table}[H]
\centering
\resizebox{1.0\textwidth}{!}{
\begin{tabular} {| l | c| c| c| c|}
\hline\hline
 \multicolumn{1}{|c|}{ Parameter} &  \multicolumn{1}{|c|}{$\Lambda$CDM} &  \multicolumn{1}{|c|}{Free-streaming DR} &  \multicolumn{1}{|c|}{Fluid DR} &  \multicolumn{1}{|c|}{Neutrinos}\\
\hline\hline
$100 \omega_b$             & $2.238~(2.253)^{+0.013}_{-0.013}   $ & $2.246~(2.248)^{+0.015}_{-0.015}   $ & $2.251~(2.252)^{+0.016}_{-0.018}   $ & $2.232~(2.23)^{+0.018}_{-0.018}   $\\
$\omega_{cdm }             $ & $0.11964~(0.11931)^{+0.00090}_{-0.00089}$ & $0.1217~(0.1202)^{+0.0012}_{-0.0022}$ & $0.1223~(0.1204)^{+0.0015}_{-0.0025}$ & $0.1183~(0.1177)^{+0.0028}_{-0.0031}$\\
$\ln 10^{10}A_s$           & $3.049~(3.053)^{+0.013}_{-0.015}   $ & $3.055~(3.049)^{+0.014}_{-0.016}   $ & $3.048~(3.048)^{+0.014}_{-0.016}   $ & $3.045~(3.051)^{+0.016}_{-0.016}   $\\
$n_{s }                    $ & $0.9652~(0.9653)^{+0.0036}_{-0.0037}$ & $0.9691~(0.9698)^{+0.0039}_{-0.0053}$ & $0.9666~(0.9656)^{+0.0038}_{-0.0038}$ & $0.9621~(0.9615)^{+0.0069}_{-0.0068}$\\
$\tau_{reio }              $ & $0.0572~(0.0578)^{+0.0067}_{-0.0075}$ & $0.0570~(0.0563)^{+0.0069}_{-0.0078}$ & $0.0577~(0.056)^{+0.0068}_{-0.0080}$ & $0.0568~(0.0597)^{+0.0067}_{-0.0075}$\\
$\Delta N_{\mbox{eff}}$    & -- & $< 0.312  $ & $< 0.285   $ & $-0.04~( -0.061)^{+0.18}_{-0.18}     $\\
$\sum m_\nu$               & $< 0.152  $ & $< 0.174                $ & $< 0.169                 $ & $< 0.146               $\\
\hline
$H_0 \,[\mathrm{km}/\mathrm{s}/\mathrm{Mpc}]$ & $67.27~(67.78)^{+0.43}_{-0.43}     $ & $67.84~(67.79)^{+0.58}_{-0.75}     $ & $68.25~(67.83)^{+0.69}_{-0.98}     $ & $66.8~(66.9)^{+1.1}_{-1.1}        $\\
$S_8$                      & $0.827~(0.826)^{+0.011}_{-0.011}   $ & $0.826~(0.8213)^{+0.012}_{-0.012}   $ & $0.826~(0.828)^{+0.011}_{-0.011}   $ & $0.826~(0.826)^{+0.011}_{-0.011}   $\\
$M_b$                      & $-19.443~(-19.43)^{+0.013}_{-0.013} $ & $-19.421~(-19.426)^{+0.015}_{-0.026} $ & $-19.412~(-19.427)^{+0.021}_{-0.030} $ & $-19.458~(-19.455)^{+0.034}_{-0.035} $\\
\hline
$H_0$ GT & $5.12\sigma $ & $4.37\sigma $ & $3.83\sigma $ & $4.19\sigma $\\
\hline
$H_0$ IT & $5.11\sigma $ & $3.53\sigma $ & $3.35\sigma $ & $4.24\sigma $\\
\hline
\end{tabular}
}
\caption{Marginalized posteriors for various model parameters for the $\Lambda$CDM, Free-streaming DR, Fluid DR, and Neutrino models, fitting to the dataset: \planck\boss\pantheon. All upper bounds are reported at 95\% C.L., for any case where the $1\sigma$ lower bound is overlapping with our priors.}
\end{table}

\begin{figure}[H]
\centering
    \includegraphics[width=0.7\textwidth]{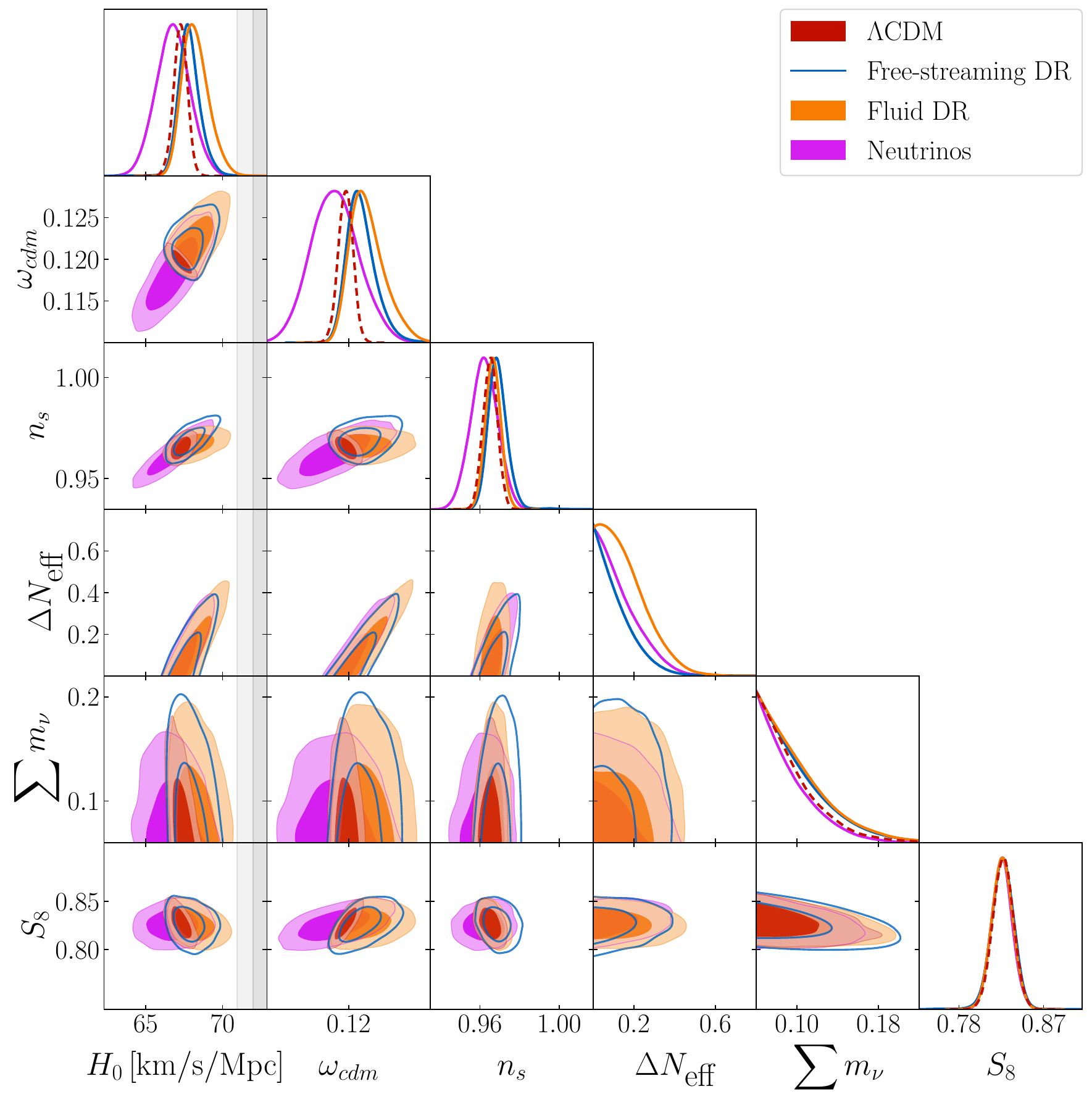}
    \caption{One and two-dimensional posterior distributions for various model parameters for the $\Lambda$CDM, Free-streaming DR, Fluid DR, and Neutrino models, fitting to the dataset: \planck\boss\pantheon. {The 68\% and 95\% confidence intervals from the measurement of $H_0$ by SH$_0$ES are shown in the gray and lighter gray shaded regions.}}
\end{figure}

\begin{table*}
\centering
\begin{tabular} {| l | c| c| c| c|}
\hline\hline
Dataset & $\Lambda$CDM & Free-streaming DR & Fluid DR & Neutrinos\\
\hline
Planck\_highl\_TTTEEE &  $2352.62$ &  $+0.82$ &  $-0.78$ &  $-2.57$\\
Planck\_lowl\_EE &  $396.83$ &  $-0.79$ &  $-0.45$ &  $+0.72$\\
Planck\_lowl\_TT &  $23.66$ &  $-0.58$ &  $-0.26$ &  $+0.48$\\
Planck\_lensing &  $8.67$ &  $+0.58$ &  $+0.11$ &  $-0.05$\\
Pantheon\_Plus &  $1411.6$ &  $-0.33$ &  $-0.38$ &  $-0.68$\\
bao\_boss\_dr12 &  $4.34$ &  $+0.19$ &  $+0.43$ &  $+0.66$\\
bao\_smallz\_2014 &  $1.24$ &  $-0.05$ &  $-0.10$ &  $-0.17$\\
Total &  $4198.96$ &  $-0.16$ &  $-1.43$ &  $-1.62$\\
\hline\hline
\end{tabular}

\caption{Values of $\chi^2$ for each likelihood when fit to a combination of \planck\boss\pantheon, reported as the difference from $\Lambda$CDM for the other models.}
\end{table*}

\subsection{Fluid Dark Radiation Produced After BBN}\label{app:FLD_afterbbn}
\begin{table}[H]
\centering
\resizebox{1.0\textwidth}{!}{
\begin{tabular} {| l | c| c| c|}
\hline\hline
 \multicolumn{1}{|c|}{ Parameter} &  \multicolumn{1}{|c|}{\planck\boss} &  \multicolumn{1}{|c|}{\planck\desi} &  \multicolumn{1}{|c|}{\planck\desi}\\
 &  \pantheon & \pantheon & \pantheon\shoes\\
\hline\hline
$100 \omega_b$             & $2.251~(2.241)^{+0.015}_{-0.017}   $ & $2.266~(2.263)^{+0.015}_{-0.019}   $ & $2.299~(2.305)^{+0.015}_{-0.015}   $\\
$\omega_{cdm }             $ & $0.1228~(0.1219)^{+0.0018}_{-0.0028}$ & $0.1229~(0.1254)^{+0.0023}_{-0.0034}$ & $0.1291~(0.1303)^{+0.0028}_{-0.0028}$\\
$\ln 10^{10}A_s$           & $3.047~(3.049)^{+0.015}_{-0.015}   $ & $3.049~(3.041)^{+0.015}_{-0.015}   $ & $3.045~(3.053)^{+0.016}_{-0.016}   $\\
$n_{s }                    $ & $0.9658~(0.9652)^{+0.0038}_{-0.0037}$ & $0.9689~(0.9666)^{+0.0037}_{-0.0037}$ & $0.9716~(0.9759)^{+0.0035}_{-0.0035}$\\
$\tau_{reio }              $ & $0.0575~(0.057)^{+0.0069}_{-0.0075}$ & $0.0607~(0.057)^{+0.0071}_{-0.0081}$ & $0.0627~(0.0679)^{+0.0073}_{-0.0083}$\\
$\Delta N_{\mbox{eff}}$    & $ < 0.433$ & $0.26~(0.34)^{+0.11}_{-0.21}      $ & $0.65~(0.73)^{+0.13}_{-0.14}      $\\
$\sum m_\nu$               & $< 0.166                 $ & $< 0.137                  $ & $< 0.149                  $\\
\hline
$H_0 \,[\mathrm{km}/\mathrm{s}/\mathrm{Mpc}]$ & $68.39~(67.94)^{+0.71}_{-1.1}      $ & $69.56~(69.82)^{+0.85}_{-1.2}      $ & $72.25~(73.0)^{+0.79}_{-0.79}     $\\
$S_8$                      & $0.826~(0.834)^{+0.011}_{-0.011}   $ & $0.815~(0.825)^{+0.010}_{-0.011}   $ & $0.809~(0.812)^{+0.011}_{-0.011}   $\\
$M_b$                      & $-19.408~(-19.42)^{+0.022}_{-0.033} $ & $-19.374~(-19.365)^{+0.026}_{-0.037} $ & $-19.298~(-19.276)^{+0.024}_{-0.021} $\\
\hline
$H_0$ GT & $3.69\sigma $ & $2.59\sigma $ & $0.6\sigma $\\
\hline
$H_0$ IT & $3.02\sigma $ & $2.28\sigma $ & $0.6\sigma $\\
\hline
$\Delta \chi^2$ & $\sim 0$ & $-0.4$ & $-24.7$\\
\hline
$\Delta$AIC & $+2.0$ & $+1.6$ & $-22.7$\\
\hline
\end{tabular}
}
\caption{Marginalized posteriors for various model parameters for the Fluid DR model where the DR is produced after BBN. The fit is shown for the datasets \planck\boss\pantheon, \planck\desi\pantheon, and \planck\desi\pantheon\shoes. All upper bounds are reported at 95\% C.L., for any case where the $1\sigma$ lower bound is overlapping with our priors.}
\end{table}

\begin{figure}[H]
\centering
    \includegraphics[width=0.7\textwidth]{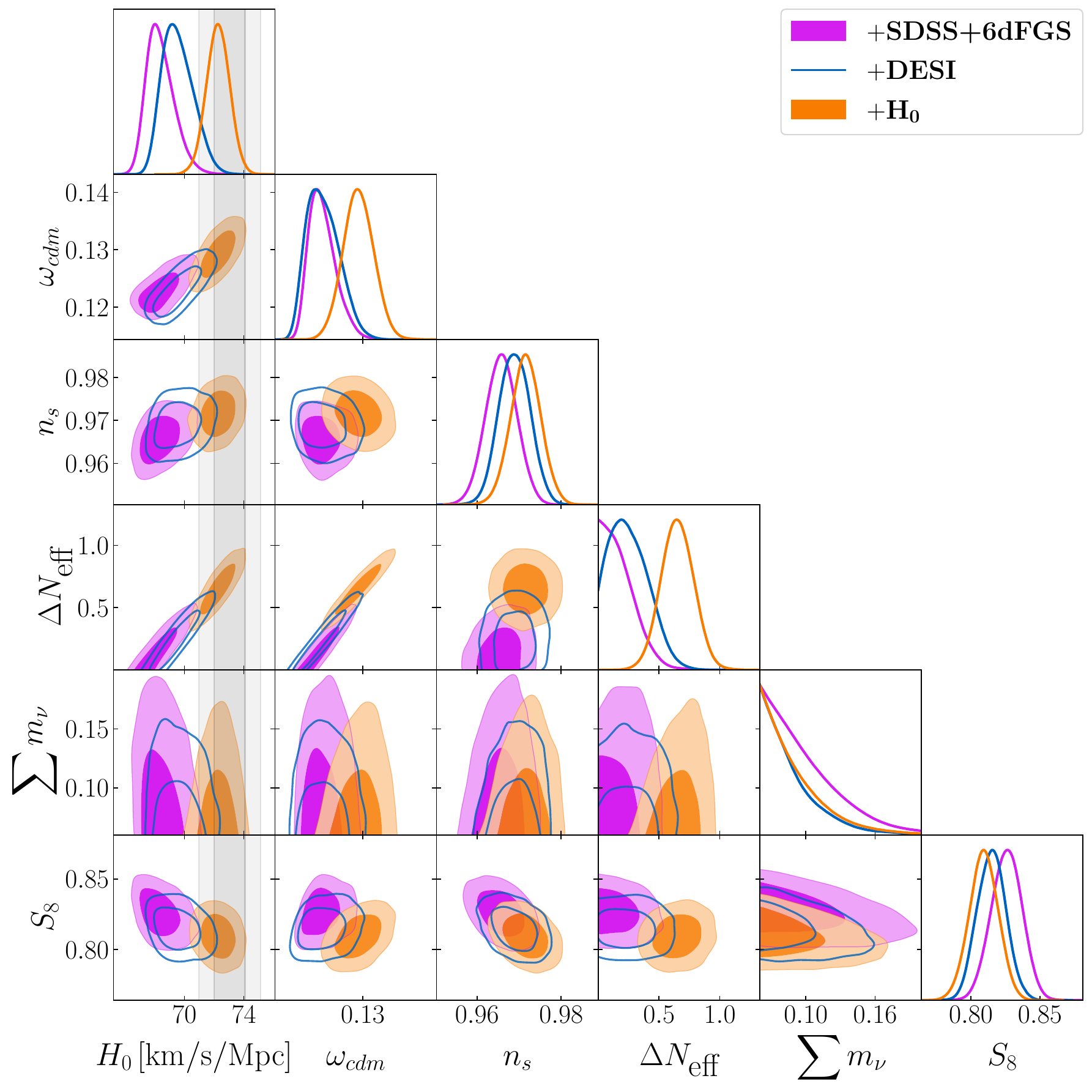}
    \caption{One and two-dimensional posterior distributions for various model parameters for the Fluid DR model where the DR is produced after BBN. The fit is shown for the datasets \planck\boss\pantheon, \planck\desi\pantheon, and \planck\desi\pantheon\shoes. {The 68\% and 95\% confidence intervals from the measurement of $H_0$ by SH$_0$ES are shown in the gray and lighter gray shaded regions.}}
\end{figure}

\begin{table*}
\resizebox{1.05\textwidth}{!}{
\begin{tabular} {| l | c| c| c|}
\hline\hline
Dataset & \planck\boss & \planck\desi & \planck\desi\\
& \pantheon & \pantheon & \pantheon\shoes\\
\hline 
Planck\_highl\_TTTEEE &  $2353.6$ &  $2354.62$ &  $2361.27$\\
Planck\_lowl\_EE &  $396.7$ &  $396.62$ &  $400.47$\\
Planck\_lowl\_TT &  $23.3$ &  $23.01$ &  $21.5$\\
Planck\_lensing &  $8.85$ &  $9.1$ &  $9.5$\\
Pantheon\_Plus &  $1411.08$ &  $1412.28$ &  --\\
Pantheon\_Plus\_shoes & -- &  -- &  $1293.09$\\
DESI\_BAO &  --  &  $16.24$ &  $13.38$\\
DESI\_BAO\_DV &  --  &  $1.27$ &  $0.24$\\
bao\_boss\_dr12 &  $5.18$ &  --  &  -- \\
bao\_smallz\_2014 &  $1.05$ &  --  &  -- \\
Total &  $4199.75$ &  $4213.13$ &  $4099.45$\\
\hline\hline
\end{tabular}
}
\caption{Values of $\chi^2$ for each likelihood for Fluid DR produced after BBN, when fit to a combination of \planck\boss\pantheon, \planck\desi\pantheon, and \planck\desi\pantheon\shoes.}
\end{table*}

\subsection{Free-streaming Dark Radiation Produced After BBN}\label{app:FS_afterbbn}

\begin{table}[H]
\centering
\resizebox{1.0\textwidth}{!}{
\begin{tabular} {| l | c| c| c|}
\hline\hline
 \multicolumn{1}{|c|}{ Parameter} &  \multicolumn{1}{|c|}{\planck\boss} &  \multicolumn{1}{|c|}{\planck\desi} &  \multicolumn{1}{|c|}{\planck\desi}\\
 & \pantheon &  \pantheon &  \pantheon\shoes\\
\hline\hline
$100 \omega_b$             & $2.245~(2.24)^{+0.015}_{-0.014}   $ & $2.257~(2.254)^{+0.015}_{-0.015}   $ & $2.288~(2.278)^{+0.014}_{-0.014}   $\\
$\omega_{cdm }             $ & $0.1219~(0.1218)^{+0.0014}_{-0.0024}$ & $0.1214~(0.1193)^{+0.0016}_{-0.0027}$ & $0.1278~(0.1287)^{+0.0026}_{-0.0026}$\\
$\ln 10^{10}A_s$           & $3.054~(3.041)^{+0.015}_{-0.016}   $ & $3.060~(3.059)^{+0.014}_{-0.017}   $ & $3.077~(3.071)^{+0.014}_{-0.017}   $\\
$n_{s }                    $ & $0.9688~(0.9676)^{+0.0043}_{-0.0050}$ & $0.9731~(0.9732)^{+0.0045}_{-0.0055}$ & $0.9864~(0.987)^{+0.0044}_{-0.0047}$\\
$\tau_{reio }              $ & $0.0568~(0.0493)^{+0.0068}_{-0.0079}$ & $0.0602~(0.0623)^{+0.0071}_{-0.0081}$ & $0.0622~(0.0584)^{+0.0069}_{-0.0084}$\\
$\Delta N_{\mbox{eff}}$    & $< 0.353$ & $< 0.435   $ & $0.63~(0.65)^{+0.14}_{-0.14}      $\\
$\sum m_\nu$               & $< 0.161  $ & $< 0.129       $ & $< 0.137           $\\
\hline
$H_0 \,[\mathrm{km}/\mathrm{s}/\mathrm{Mpc}]$ & $68.03~(68.17)^{+0.57}_{-0.84}     $ & $68.94~(68.41)^{+0.63}_{-0.99}     $ & $71.82~(71.65)^{+0.78}_{-0.77}     $\\
$S_8$                      & $0.830~(0.826)^{+0.011}_{-0.011}   $ & $0.821~(0.822)^{+0.011}_{-0.011}   $ & $0.823~(0.83)^{+0.011}_{-0.011}   $\\
$M_b$                      & $-19.419~(-19.414)^{+0.017}_{-0.026} $ & $-19.393~(-19.41)^{+0.019}_{-0.030} $ & $-19.310~(-19.311)^{+0.022}_{-0.022} $\\
\hline
$H_0$ GT & $4.22\sigma $ & $3.37\sigma $ & $0.94\sigma $\\
\hline
$H_0$ IT & $3.62\sigma $ & $2.84\sigma $ & $0.94\sigma $\\
\hline
$\Delta \chi^2$ & $\sim 0$ & $+0.4$ & $-20.5$\\
\hline
$\Delta$AIC & $+2.0$ & $+2.4$ & $-18.5$\\
\hline
\end{tabular}
}
\caption{Marginalized posteriors for various model parameters for the free-streaming DR model where the DR is produced after BBN. The fit is shown for the datasets \planck\boss\pantheon, \planck\desi\pantheon, and \planck\desi\pantheon\shoes. All upper bounds are reported at 95\% C.L., for any case where the $1\sigma$ lower bound is overlapping with our priors.}
\end{table}

\begin{figure}[H]
\centering
    \includegraphics[width=0.7\textwidth]{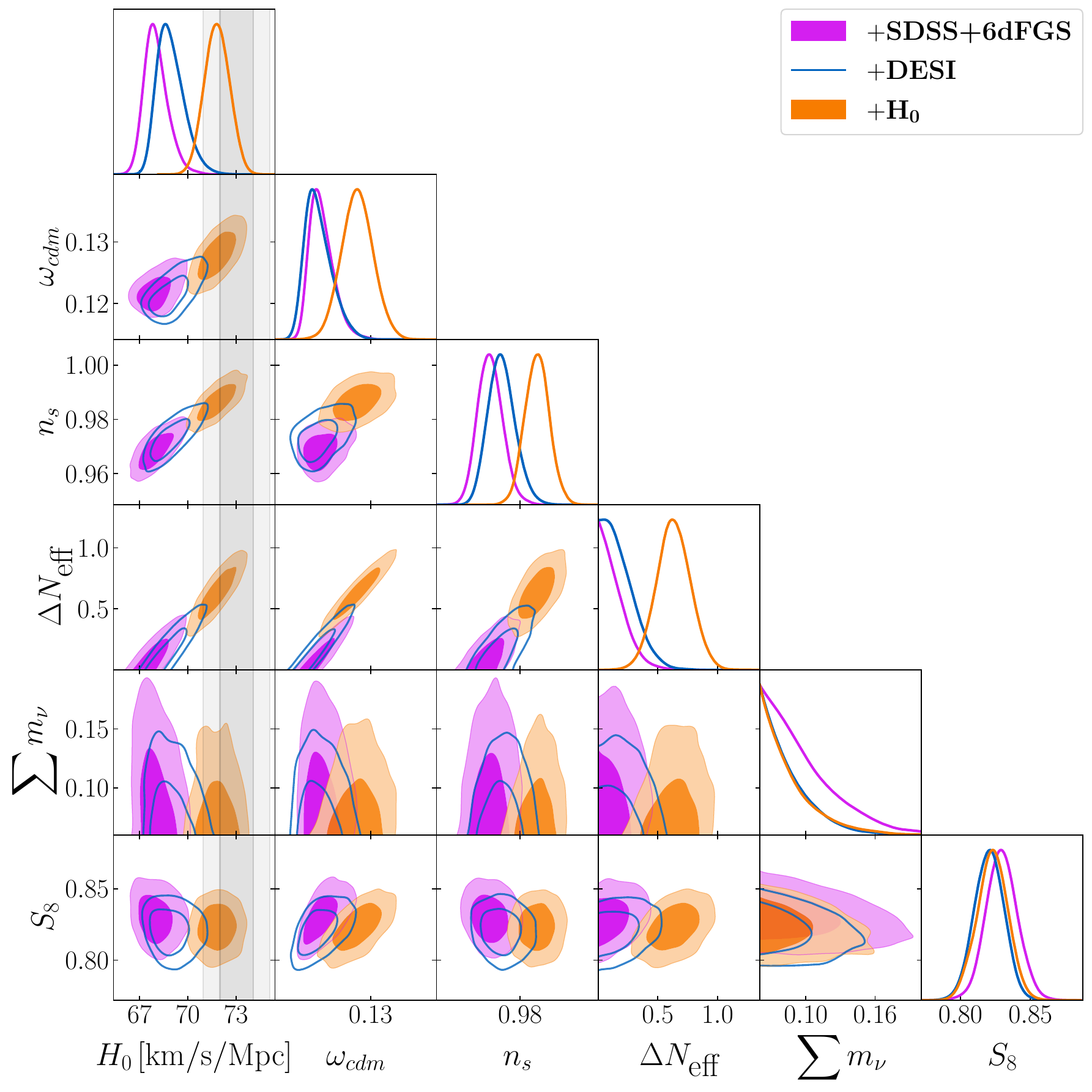}
    \caption{One and two-dimensional posterior distributions for various model parameters for the free-streaming DR model where the DR is produced after BBN. The fit is shown for the datasets \planck\boss\pantheon, \planck\desi\pantheon, and \planck\desi\pantheon\shoes. {The 68\% and 95\% confidence intervals from the measurement of $H_0$ by SH$_0$ES are shown in the gray and lighter gray shaded regions.}}
\end{figure}

\begin{table*}
\begin{tabular} {| l | c| c| c|}
\hline\hline
Dataset & \planck\boss & \planck\desi & \planck\desi\\
 & \pantheon & \pantheon & \pantheon\shoes\\
\hline
Planck\_highl\_TTTEEE &  $2355.38$ &  $2355.14$ &  $2366.39$\\
Planck\_lowl\_EE &  $395.72$ &  $398.1$ &  $396.73$\\
Planck\_lowl\_TT &  $22.97$ &  $22.2$ &  $21.11$\\
Planck\_lensing &  $9.45$ &  $8.97$ &  $9.7$\\
Pantheon\_Plus &  $1411.23$ &  $1412.37$ &  --\\
Pantheon\_Plus\_shoes &  -- &  -- &  $1296.27$\\
DESI\_BAO &  --  &  $16.1$ &  $14.35$\\
DESI\_BAO\_DV &  --  &  $1.22$ &  $0.61$\\
bao\_boss\_dr12 &  $4.72$ &  --  &  -- \\
bao\_smallz\_2014 &  $1.15$ &  --  &  -- \\
Total &  $4200.62$ &  $4214.11$ &  $4105.15$\\
\hline\hline
\end{tabular}
\caption{Values of $\chi^2$ for each likelihood for FS DR produced after BBN, when fit to a combination of \planck\boss\pantheon, \planck\desi\pantheon, and \planck\desi\pantheon\shoes.}
\end{table*}

\subsection{Fluid Dark Radiation Produced After BBN with \planckhilli~ and \DES}\label{app:planckhilli_DES}

\begin{table}[H]
\centering

\begin{tabular} {| l | c| c| c|}
\hline\hline
 \multicolumn{1}{|c|}{ Parameter} &  \multicolumn{1}{|c|}{\planckhilli\desi} &  \multicolumn{1}{|c|}{\planckhilli\desi} &  \multicolumn{1}{|c|}{\planck\desi}\\
 &  \pantheon &  \DES & \DES\\
\hline\hline
$100 \omega_b$             & $2.254~(2.256)^{+0.017}_{-0.017}   $ & $2.249~(2.253)^{+0.016}_{-0.017}   $ & $2.260~(2.245)^{+0.016}_{-0.018}   $\\
$\omega_{cdm }             $ & $0.1239~(0.1228)^{+0.0028}_{-0.0034}$ & $0.1235~(0.1238)^{+0.0025}_{-0.0035}$ & $0.1228~(0.1188)^{+0.0020}_{-0.0034}$\\
$\ln 10^{10}A_s$           & $3.046~(3.041)^{+0.013}_{-0.013}   $ & $3.046~(3.049)^{+0.013}_{-0.013}   $ & $3.048~(3.056)^{+0.014}_{-0.016}   $\\
$n_{s }                    $ & $0.9699~(0.9707)^{+0.0037}_{-0.0036}$ & $0.9690~(0.9683)^{+0.0037}_{-0.0037}$ & $0.9679~(0.968)^{+0.0038}_{-0.0038}$\\
$\tau_{reio }              $ & $0.0624~(0.061)^{+0.0063}_{-0.0063}$ & $0.0618~(0.0644)^{+0.0062}_{-0.0062}$ & $0.0595~(0.0622)^{+0.0068}_{-0.0081}$\\
$\Delta N_{\mbox{eff}}$    & $0.35~(0.33)^{+0.16}_{-0.20}      $ & $0.31~(0.31)^{+0.13}_{-0.21}      $ & $0.231~(0.019)^{+0.062}_{-0.22}    $\\
$\sum m_\nu$               & $< 0.172                   $ & $< 0.181                   $ & $< 0.152                  $\\
\hline
$H_0 \,[\mathrm{km}/\mathrm{s}/\mathrm{Mpc}]$ & $70.0~(70.5)^{+1.1}_{-1.2}        $ & $69.54~(69.78)^{+0.92}_{-1.2}      $ & $69.13~(68.07)^{+0.79}_{-1.2}      $\\
$S_8$                      & $0.810~(0.802)^{+0.010}_{-0.010}   $ & $0.813~(0.818)^{+0.011}_{-0.0099}  $ & $0.819~(0.82)^{+0.010}_{-0.011}   $\\
$M_b$                      & $-19.360~(-19.35)^{+0.032}_{-0.037} $ &                              &                             \\
\hline
$H_0$ GT & $2.03\sigma $ & $2.52\sigma $ & $2.99\sigma $\\
\hline
$H_0$ IT & $1.89\sigma $ & $2.21\sigma $ & $2.51\sigma $\\
\hline
$\Delta \chi^2$ & $-1.4$ & $-0.6$ & $-0.2$\\
\hline
$\Delta$AIC & $+0.6$ & $+1.4$ & $+1.8$\\
\hline
\end{tabular}
\caption{Marginalized posteriors for various model parameters for the Fluid DR model where the DR is produced after BBN. The fit is shown for the datasets \planckhilli\desi\pantheon and \planckhilli\desi\DES. All upper bounds are reported at 95\% C.L., for any case where the $1\sigma$ lower bound is overlapping with our priors.}
\end{table}

\begin{figure}[H]
\centering
    \includegraphics[width=0.7\textwidth]{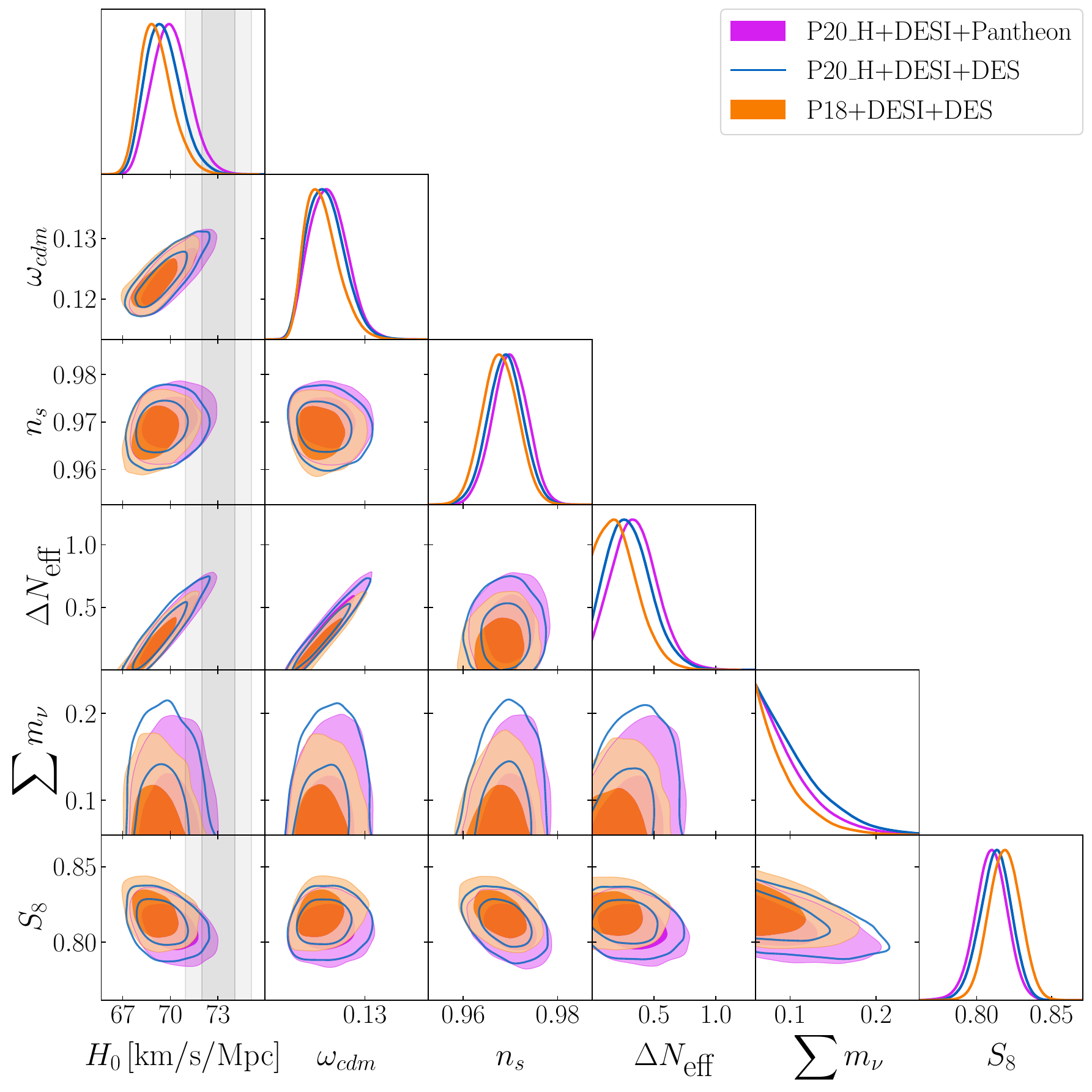}
    \caption{One and two-dimensional posterior distributions for various model parameters for the Fluid DR model where the DR is produced after BBN. The fit is shown for the datasets \planckhilli\desi\pantheon, \planckhilli\desi\DES, and \planck\desi\DES. {The 68\% and 95\% confidence intervals from the measurement of $H_0$ by SH$_0$ES are shown in the gray and lighter gray shaded regions.}}
\end{figure}

\begin{table*}
\begin{tabular} {| l | c| c| c|}
\hline\hline
Dataset & \planckhilli\desi & \planckhilli\desi & \planck\desi\\
 & \pantheon & \DES & \DES\\
\hline
Planck\_highl\_TTTEEE &  --  &  --  &  $2351.1$\\
Planck\_lowl\_EE &  --  &  --  &  $398.25$\\
Planck\_lowl\_TT &  $22.11$ &  $22.65$ &  $23.04$\\
Planck\_lensing &  $11.39$ &  $9.61$ &  $8.91$\\
Planck20\_Hillipop\_TTTEEE &  $30512.4$ &  $30513.35$ &  -- \\
Planck20\_Lollipop\_EE &  $33.01$ &  $33.87$ &  -- \\
Pantheon\_Plus &  $1414.93$ &  --  &  -- \\
DES &  --  &  $1649.74$ &  $1648.86$\\
DESI\_BAO &  $13.42$ &  $15.04$ &  $16.74$\\
DESI\_BAO\_DV &  $0.37$ &  $0.93$ &  $1.42$\\
bao\_boss\_dr12 &  --  &  --  &  -- \\
bao\_smallz\_2014 &  --  &  --  &  -- \\
Total &  $32007.64$ &  $32245.18$ &  $4448.33$\\
\hline\hline
\end{tabular}
\caption{Values of $\chi^2$ for each likelihood for the Fluid DR produced after BBN, when fit to a combination of \planckhilli\desi\pantheon, \planckhilli\desi\DES, and \planck\desi\DES.}
\end{table*}

\end{document}